\documentclass[a4paper,superscriptaddress,nofootinbib]{revtex4-1}

\usepackage[normalem]{ulem}

\usepackage{makecell}

\usepackage{epsfig}
\usepackage{amsfonts}
\usepackage{amsmath}
\usepackage{amssymb}
\usepackage{amsthm}
\usepackage{array}
\usepackage{booktabs}
\usepackage{color}
\usepackage{graphicx}
\usepackage{hyperref}
\usepackage{multirow}
\usepackage{slashed}
\usepackage{subfigure}
\usepackage{theorem}
\usepackage{textcomp}

\allowdisplaybreaks[1]

\numberwithin{equation}{section}

\begin{document}

\title{Constraining $Z^\prime$ widths from $p_T$ measurements in Drell-Yan processes}

\author{Elena Accomando}
\email[E-mail: ]{e.accomando@soton.ac.uk}
\affiliation{School of Physics \& Astronomy, University of Southampton,
        Highfield, Southampton SO17 1BJ, UK}
\affiliation{Particle Physics Department, Rutherford Appleton Laboratory, 
       Chilton, Didcot, Oxon OX11 0QX, UK}

\author{Juri Fiaschi}
\email[E-mail: ]{juri.fiaschi@soton.ac.uk}
\affiliation{School of Physics \& Astronomy, University of Southampton,
        Highfield, Southampton SO17 1BJ, UK}
\affiliation{Particle Physics Department, Rutherford Appleton Laboratory, 
       Chilton, Didcot, Oxon OX11 0QX, UK}

\author{Stefano Moretti}
\email[E-mail: ]{s.moretti@soton.ac.uk}
\affiliation{School of Physics \& Astronomy, University of Southampton,
        Highfield, Southampton SO17 1BJ, UK}
\affiliation{Particle Physics Department, Rutherford Appleton Laboratory, 
       Chilton, Didcot, Oxon OX11 0QX, UK}

\author{Claire H. Shepherd-Themistocleous}
\email[E-mail: ]{claire.shepherd@stfc.ac.uk}
\affiliation{School of Physics \& Astronomy, University of Southampton,
        Highfield, Southampton SO17 1BJ, UK}
\affiliation{Particle Physics Department, Rutherford Appleton Laboratory, 
       Chilton, Didcot, Oxon OX11 0QX, UK}

\vspace*{1.0truecm}
\begin{abstract}
{
\centerline{\bf Abstract}

We define a Focus Point (FP) Asymmetry, $A_{\rm FP}$, obtained by integrating the normalised transverse momentum distribution of either lepton produced in the Drell-Yan (DY) process below and above a point where a variety of popular $Z^\prime$ models all have the same magnitude. 
For a given $Z^\prime$ mass the position of this FP is predictable, depending only on the collider energy and on the low transverse momentum cut chosen in the normalisation procedure. 
The resulting $A_{\rm FP}$ is very sensitive to the $Z^\prime$ width and can be used to constrain this parameter in experimental fits.
}

\end{abstract}

\maketitle

\setcounter{footnote}{0}
\setcounter{page}{0}

\section{Introduction}

Additional massive neutral gauge bosons, also known as $Z^\prime$s,
are ubiquitous in Beyond Standard Model (BSM) scenarios.
Experimentally such states can be observed in invariant mass spectra
formed using the decay products of the $Z^\prime$ in for example a
di-lepton mass spectrum. The new physics signal has some peaking
structure, concentrated in some interval centred around its
mass. Experimental searches for such heavy states often assume that
such a resonance can be described by a Breit-Wigner (BW) line-shape,
above a smooth SM background.

A $Z^\prime$ resonance can have a wide range of intrinsic widths,
which depend on the scenario considered. It can be narrow, as for
example, in $E_6$, Generalised Left-Right (GLR) symmetric and
Generalised Standard Model (GSM) scenarios \cite{Accomando:2010fz},
where $\Gamma_{Z^\prime} / M_{Z^\prime} \sim 0.5 - 10
\%$. Alternatively, it can be wide, as in Technicolour
\cite{Belyaev:2008yj} scenarios, Composite Higgs Models
\cite{Barducci:2012kk} or in more generic models where the $Z^\prime$
boson coupling to the first two fermion generations is different to
that the the third generation \cite{Kim:2014afa, Malkawi:1999sa}. The
$Z^\prime$ can also interact with the SM gauge bosons in presence of
$Z/Z^\prime$ mixing \cite{Altarelli:1989ff}. In all of these cases
large $\Gamma_{Z^\prime}/M_{Z^\prime}$ values, up to $\sim 50 \%$, are
induced by the additional $Z^\prime$ decay channels available in all
such cases. When very wide the resonance does not have a well-defined
BW line-shape and appears as a broad shoulder over the SM background.

The most generic experimental analyses look for narrow resonances
where the experimental resolution is the dominant contribution to the
observable width of a peak structure appearing over a SM background.
In this approach, theoretical cross section predictions for specific
models are usually calculated in the Narrow Width Approximation (NWA).
Finite Width (FW) and interference effects can be taken into account
in a model independent way following the approach described in
\cite{Accomando:2013sfa}.
Up to date experimental bounds on narrow (i.e., where $\Gamma_{Z^\prime}/M_{Z^\prime}\sim 1\%$) $Z^\prime$  
resonances have been released from CMS~\cite{CMS:2016abv}
and ATLAS~\cite{ATLAS:2017wce} with the Run 2  energy of 13 TeV
and an integrated luminosity of 13 fb$^{-1}$ and 36.1 fb$^{-1}$ respectively.
The most stringent bounds set the limit for the masses of these objects $M_{Z^\prime} > 4$ TeV.
For wider $Z^\prime$s, the experimental collaborations look for both resonances and
effectively very wide resonances in non-resonant searches. In the first case, 
ATLAS has provided us with acceptance curves that can be used 
to rescale the limits obtained for narrow resonances, for widths up to 5--10\% 
of the mass at the most~\cite{ATLAS:2017wce}.
In the second (`effectively' non-resonant case, where the width-to-mass region can be over 10\%), the experimental analyses are
essentially counting experiments: an excess of events is searched for
above an estimated SM background. These last searches optimize selection
criteria in the context of particular specific models order to
maximise the discovery/exclusion potential at the LHC. The
experimental results heavily rely on the good understanding and
control of the SM background. In this case, the use of charge
asymmetries may be useful in extracting a $Z^\prime$ signal
\cite{Accomando:2015cfa}. (Needless to say, in the remainder, we will define benchmarks which escapes experimental limits, for any value
of $\Gamma_{Z^\prime}/M_{Z^\prime}$ presented.)

If a $Z^\prime$ state were to be observed at the LHC determining the
intrinsic width would be an immediate objective. The width would
provide information about the underlying $Z^\prime$ model and the
coupling strength and quantum numbers of the $Z^\prime$ in its
interactions with SM objects. The measurement of a width using the
mass spectrum is limited by the detector resolution in the case of a
narrow resonance and for a very wide resonance (that cannot be
approximated by a BW) a model specific approach would be required.

The purpose of this paper is to describe the role of an alternative
observable to the di-lepton invariant mass ($M_{ll}$) that could be
used to extract information on the intrinsic width of the $Z^\prime$.
The advantage of this approach is twofold. Firstly, one can use this
new observable to determine the intrinsic width of the
resonance. Secondly, the latter can potentially be used to perform a
constrained fit to the cross section (or charge asymmetry) in the
di-lepton invariant mass, so as to disentangle the pure signal
contributions from dynamics resulting from FW and/or interference
effects. (While we will address the first point in this publication,
we will defer treatment of the second to a forthcoming one.) This new
observable is the transverse momentum distribution of an individual
lepton in the final state. We will show that the corresponding
(normalised) spectrum  exhibits a Focus Point (FP) that is the same
for all $Z^\prime$ models considered, the latter thereby acting
similarly to the $Z^\prime$ pole in the di-lepton invariant mass. One
can also define asymmetries around this FP, $A_{\rm FP}$s,
that provide information
on the underlying $Z^\prime$ scenario, in terms of its quantum
numbers. 

This is in principle analogous to the case of charge asymmetries, in practice though the FP ones  display sensitivity to a different parameter.
In fact, herein, we assume that a $Z^\prime$ state has already been observed and a (tentative) value
of its mass has been extracted: this is a precondition to the exploitation of the FP and its asymmetries. With this mind,
such FP observables provide one with an additional powerful diagnostic tool in understanding the nature of the $Z^\prime$, quite uncorrelated
to the aforementioned cross section and  charge asymmetries, as they display a strong sensitivity to its width, whichever the actual value of it.
This is extremely important as, on the one hand, $\Gamma_{Z^\prime}$ contains information about all couplings of the $Z^\prime$ state (hence about
the underlying model) and, on the other hand, neither fits to  the cross section (wherein the dependence upon $\Gamma_{Z^\prime}$ really ought to be
minimized in the search for the BW peak)  nor mappings of charge asymmetries (which are primarily sensitive to the relative sign of the above couplings) offer the
same scope\footnote{We also remark here that the use of an $A_{\rm FP}$ as a search variable of a $Z^\prime$ state, along similar lines to those put forward
for, e.g., $A_{\rm FB}$ \cite{Accomando:2015cfa}, can also be conceived, though this is beyond the remit of this paper.}.

This note is organized as follows. In Sect.~\ref{sec:pT} we
introduce the new variable and describe how it can be used for the
aforementioned purposes. In Sect.~\ref{sec:width} we illustrate our
results. Finally, we conclude in Sect.~\ref{sec:summa}.

\section{$Z^\prime$s $p_T$ distribution spectra}
\label{sec:pT}

In order to perform our analysis we have used the numerical code
documented in Refs.~\cite{Accomando:2013sfa,Accomando:2015cfa}.
Standard acceptance cuts on the leptons have been required: $p_T>20$
GeV and $|\eta|<2.5$. The acceptance $p_T$ cut is not really important in our analysis since we are going to introduce 
a substantial $p_T^{\rm min}$ cut on the leptons ($>~900$ GeV) when analysing our transverse momentum distribution.
Moreover we have verified that tightening the pseudorapidity does not change our
conclusions, as discussed in Sect.~\ref{subsec:eta_cut}.
In order to speed up the numerical simulation (we will be working with very high invariant
masses, of ${\cal O}(1~{\rm TeV})$, we require that $M_{ll}>50$ GeV.

Differential distributions for three $Z^\prime$ benchmark models
($E_6^I$, GLR-LR, GSM-SSM \cite{Accomando:2015cfa}) have been
generated for different $Z^\prime$ boson masses and widths\footnote{These models has been chosen as representative of their 
respective class, since we will show in Sect.\ref{subsec:sensitivity} that the new analysis 
will produce similar results for all single $Z^\prime$ models therein.}.
In computing the binned number of events, we include all the contributions to the same final state:
$Z^\prime$ signal, SM background and their mutual interference.
Higher orders corrections have not been considered in this work.
Both NNLO QCD and NLO EW corrections can be large, but they also contribute with opposite signs, leading to some cancellations~\cite{Balossini:2008cs}.
However we are interested in the very high $p_T$ region, where we can assume the NNLO QCD contribution to appear as a (roughly) constant k-factor~\cite{Li:2012wna}.
The asymmetry observable that we will define in the following will naturally provide a cancellation of this effect.
NLO EW corrections instead are expected to grow in magnitude with the energy, and they might lead to observable effects.
Yet, no public code is available at the moment for the NLO calculation of EW radiative corrections to the leptons' $p_T$ spectra in DY production
including real and virtual EW gauge boson emission, both of which are needed for an accurate estimate of the effects we are studying,
owing to the fact that the di-lepton final state is treated inclusively in experimental analyses
(i.e., no veto is enforced against real radiation of EW gauge bosons). Hence, for the time being, we will neglect these effects too.

\begin{figure}[h]
\begin{center}
\includegraphics[width=0.45\textwidth]{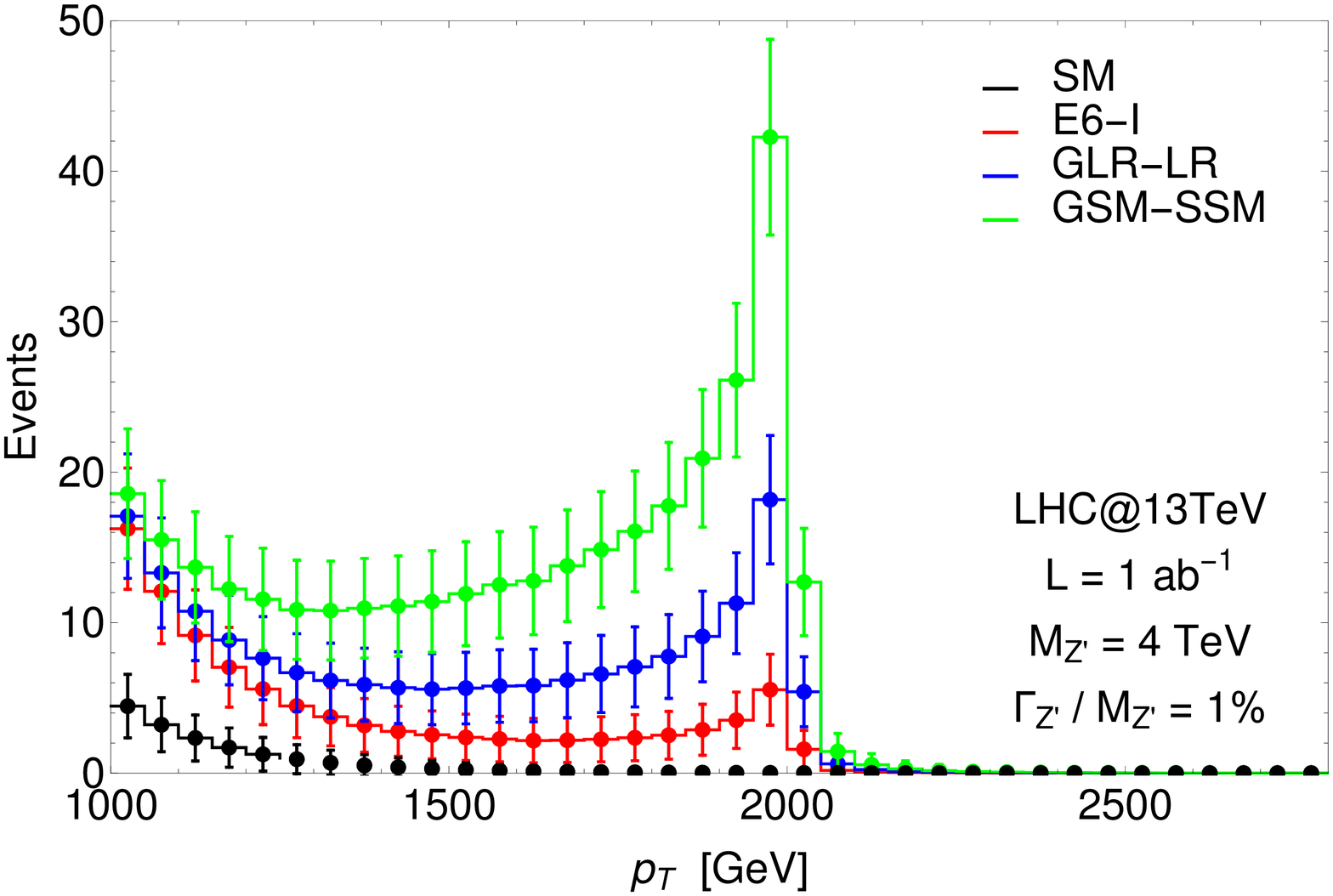}{(a)}
\includegraphics[width=0.45\textwidth]{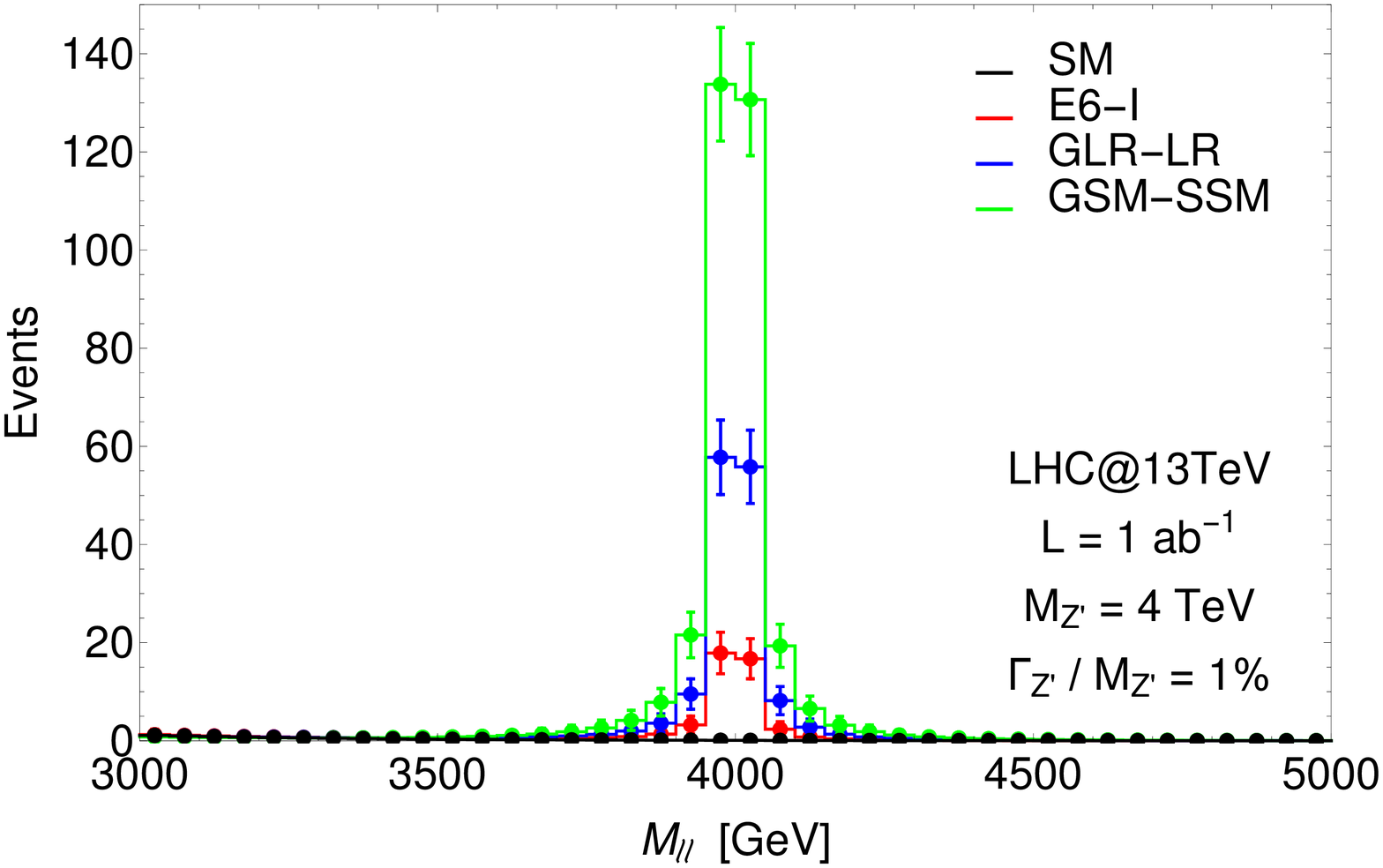}{(b)}
\caption{Distribution of number of events as function of (a) the $p_T$ of either lepton and (b) of the di-lepton invariant mass
as predicted in the SM and in three $Z^\prime$ benchmark models with $M_{Z^\prime}=4~{\rm TeV}$ at the 13 TeV LHC with $\mathcal{L}=1~ab^{-1}$.
For all models the width of the resonance has been fixed at 1\% of its mass.
Acceptance cuts are applied ($|\eta|<2.5$), detector efficiencies are not accounted for.}
\label{fig:pT_models}
\end{center}
\end{figure}

In Fig.~\ref{fig:pT_models} we show the $p_T$ and the invariant mass
distributions. The data shown have been binned by integrating in the
$p_T$ ($M_{ll}$) variable and multiplying by the quoted luminosity in
order to obtain the number of events on the $y$ axis. The error bars
represent the statistical error on the number of events observed in
each bin and are given by the square root of the number of events in
each bin. As expected in the $p_T$ distribution, a noticeable peak
appears at $p_T\approx M_{Z^\prime}/2$ for all BSM scenarios
considered with the slope leading to it varying depending on the
underlying $Z^\prime$ model. The total number of events is defined by
the model cross section. The SM distribution by contrast monotonically
decreases. There is no point in $p_T$ amongst the various curves
where all the differential cross sections have the same magnitude.

\begin{figure}[h]
\begin{center}
\includegraphics[width=0.45\textwidth]{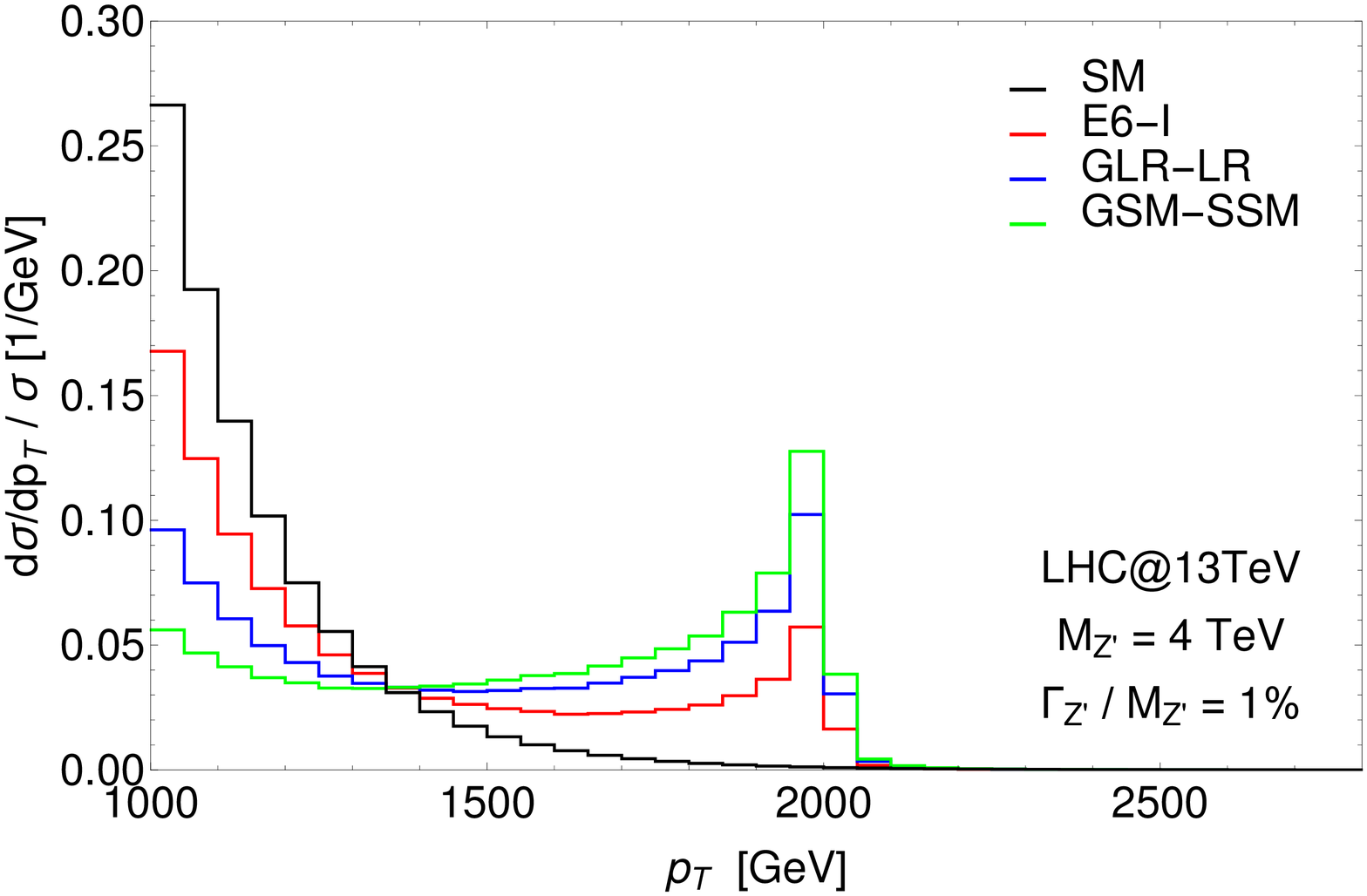}{(a)}
\includegraphics[width=0.45\textwidth]{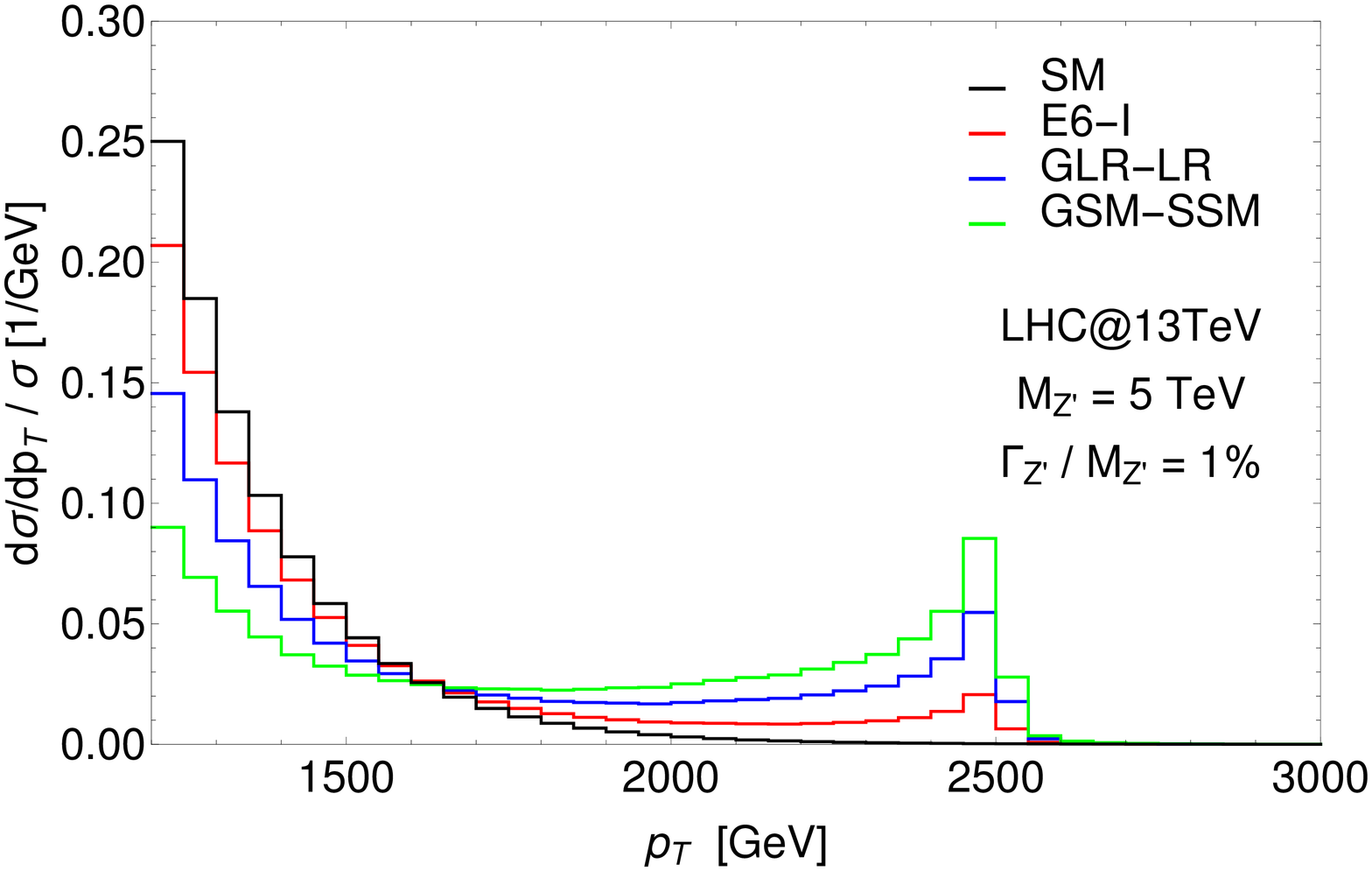}{(b)}
\caption{Normalised distribution in $p_T$ of either lepton as predicted in the SM and in three $Z^\prime$ benchmark models at the 13 TeV LHC.
For all models the width of the resonance has been fixed at 1\% of its mass.
Acceptance cuts are applied ($|\eta|<2.5$), detector efficiencies are not accounted for.
(a) $p_T^{\rm min} = 1000~{\rm GeV}$ and $M_{Z^\prime}=4~{\rm TeV}$, (b) $p_T^{\rm min} = 1200~{\rm GeV}$ and $M_{Z^\prime}=5~{\rm TeV}$.
}
\label{fig:pT_norm}
\end{center}
\end{figure}

An interesting feature appears when the distributions are normalised.
Starting from the differential
distributions shown in Fig. \ref{fig:pT_models}(a) for each model in the legend,
we divide the number of events in each bin by the total number of events that is obtained integrating 
the cross section from the chosen $p_T^{\rm min}$ on. For this specific case we chose $p_T^{\rm min} = 1000$ GeV.
The results of this normalisation are shown in Fig.~\ref{fig:pT_norm}(a). The most
interesting feature in this plot is that around $p_T$ = 1400 GeV all
the curves have the same magnitude. We call this intersection point the Focus Point (FP). 
The FP position strongly depends on the lepton $p_T^{\rm min}$ cut that we choose to maximise the sensitivity 
to the hypothetical $Z^\prime$ boson. This will be discussed more extensively in Sect.~\ref{subsec:sensitivity}, 
here we give just an example of this effect. For a 
$Z^\prime$ mass of 5 TeV the optimal choice is $p_T^{\rm min} = 1200$ GeV. In this case we obtain very similar
behaviour, albeit with the FP shifted to around 1.2 TeV, as plotted in Fig.~\ref{fig:pT_norm}(b). In these
illustrations we have taken the LHC energy to be 13 TeV and use the
CT14NNLO PDF set \cite{Dulat:2015mca} evaluated at the $Q=\sqrt{\hat s}$
factorisation/renormalisation scale (i.e., the centre-of-mass energy at the parton level).

\begin{figure}[h]
\begin{center}
\includegraphics[width=0.45\textwidth]{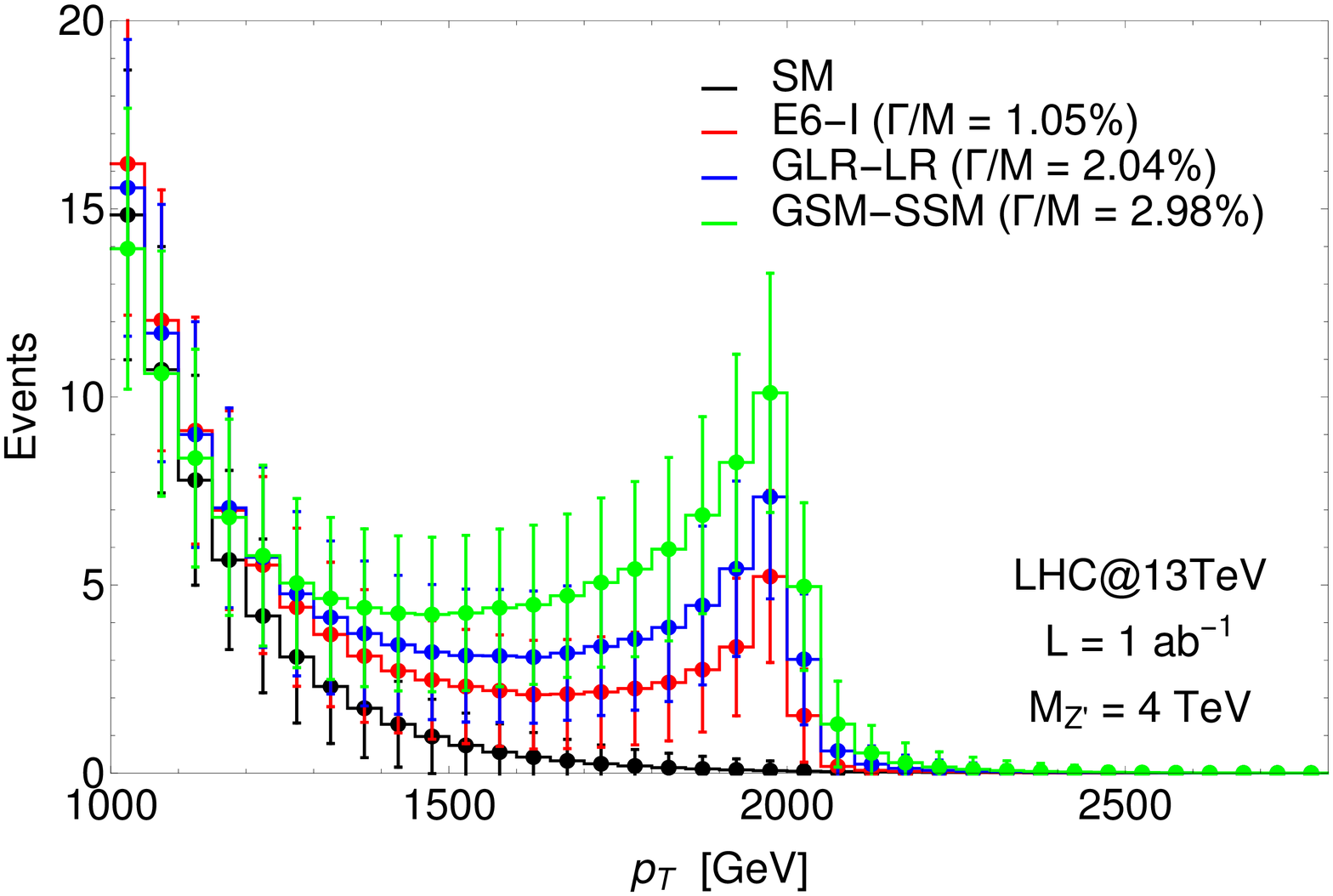}{(a)}
\includegraphics[width=0.45\textwidth]{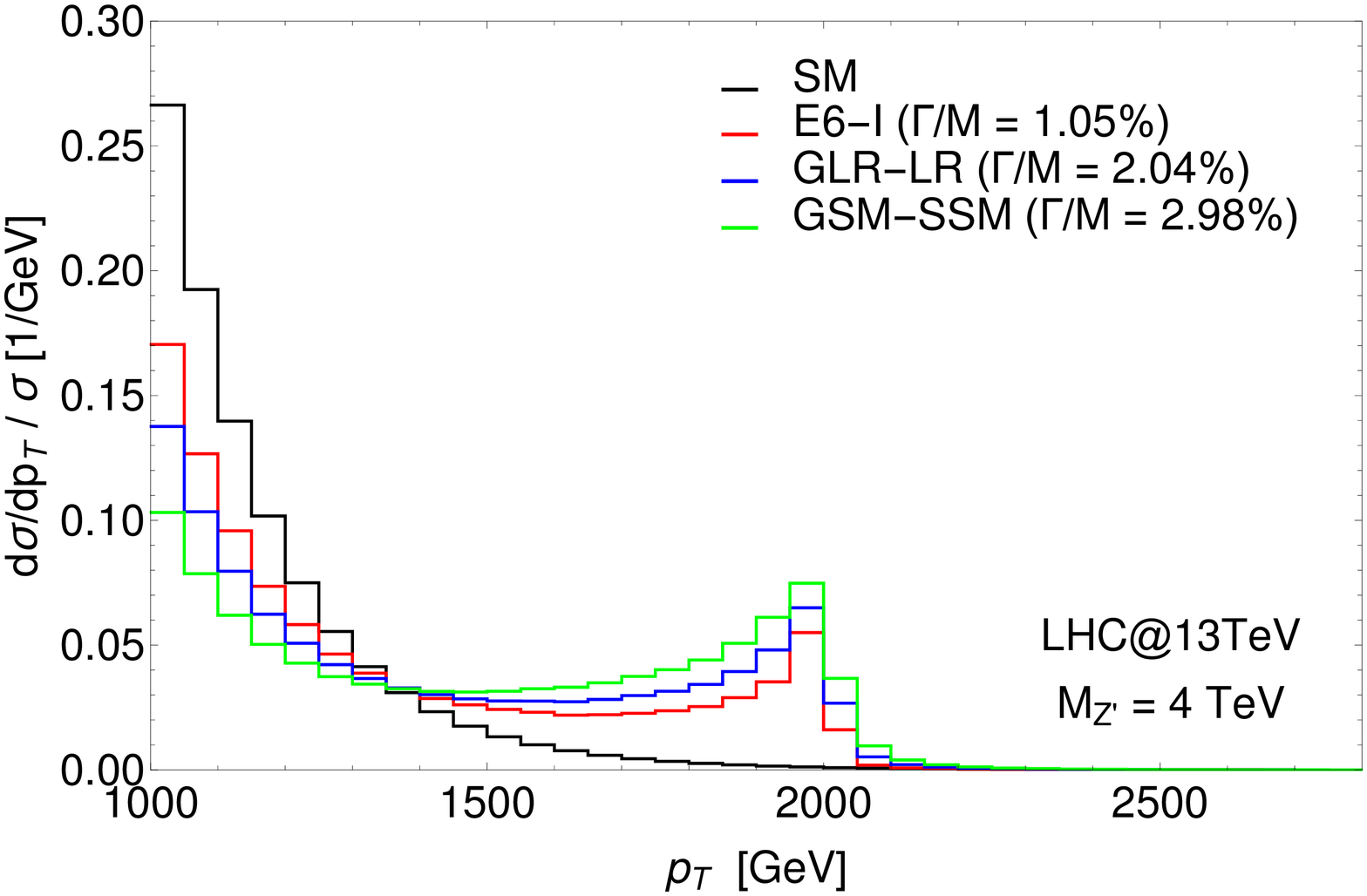}{(b)}
\caption{(a) Number of events as function of $p_T$ of either lepton 
as predicted in the SM and in three $Z^\prime$ benchmark models with $M_{Z^\prime}=4~{\rm TeV}$ at the 13 TeV LHC with $\mathcal{L}=1~ab^{-1}$.
The width of the resonances has been fixed at their natural value as predicted by the model.
Acceptance cuts are applied ($|\eta|<2.5$), no detector efficiencies are accounted for.
(b) Normalized distribution of (a) with $p_T^{\rm min} = 1000~GeV$.}
\label{fig:pT_natural_width}
\end{center}
\end{figure}

For completeness in Fig.~\ref{fig:pT_natural_width} we show
distributions for the number of events and the normalized $p_T$ for
the three benchmark models with the resonance widths fixed to the
natural values predicted by each model. The values for the resonance
widths can be significantly modified by the presence of new physics,
therefore in order to be as general as possible we will consider the
$Z^\prime$ width to be a free parameter.

In order to understand this feature in detail, in the following
section we explore its dependence upon the collider energy, the
$Z^\prime$ parameters (its mass and width), the minimum $p_T$ cut and
the normalisation procedure as well as the role of the interference
between the $Z^\prime$ diagram and SM topologies. By contrast, we
limit ourselves to simply state here that we have verified the
independence of the FP location upon the choice of PDFs and $Q$: this
should not be surprising as the quark and antiquark behaviour inside
the proton at the relevant $x$ and $Q$ values is well known
\cite{Accomando:2016tah}.

\subsection{The role of the partonic (or collider) energy}

\begin{figure}[h]
\begin{center}
\includegraphics[width=0.45\textwidth]{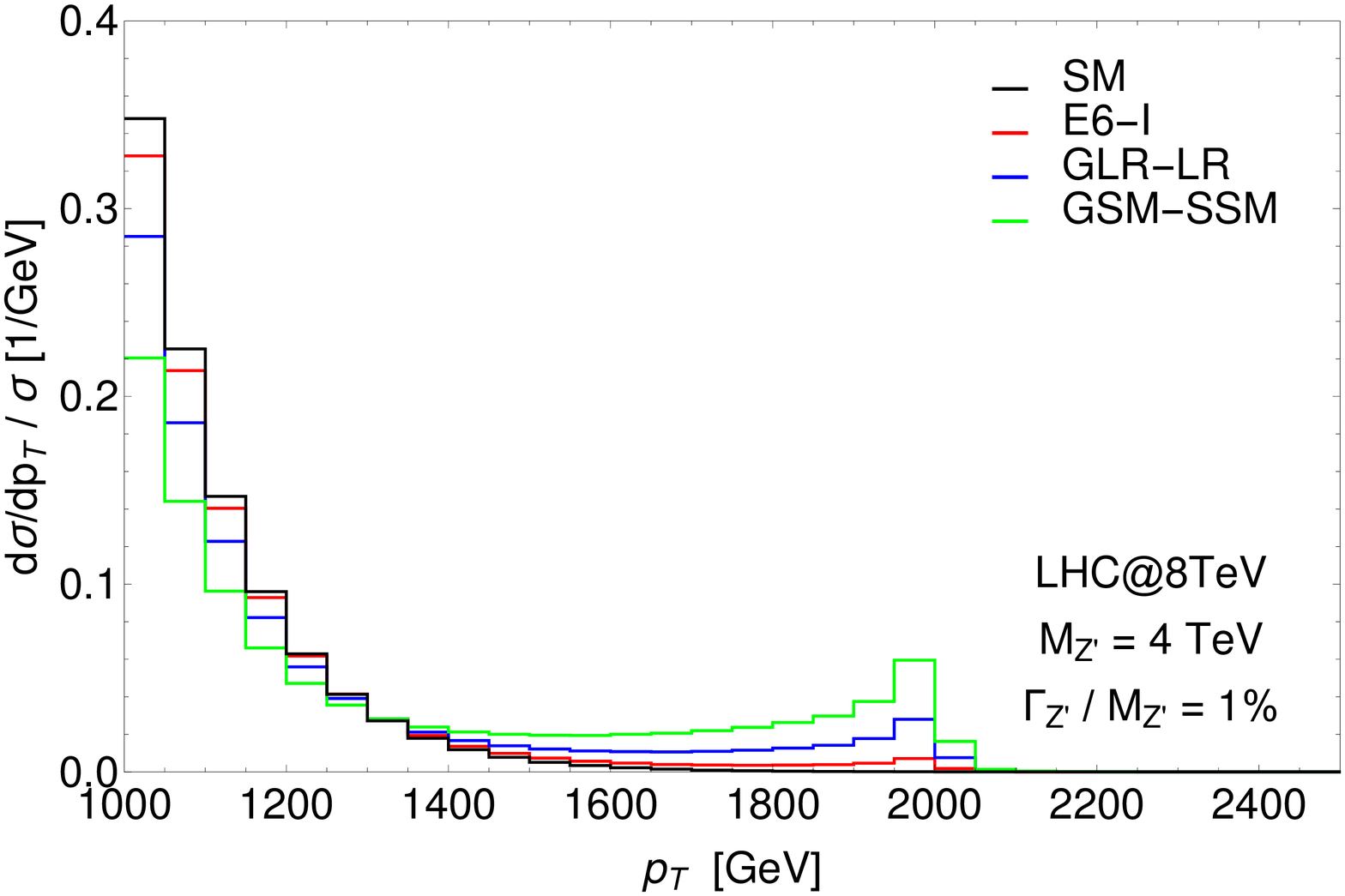}{(a)}
\includegraphics[width=0.45\textwidth]{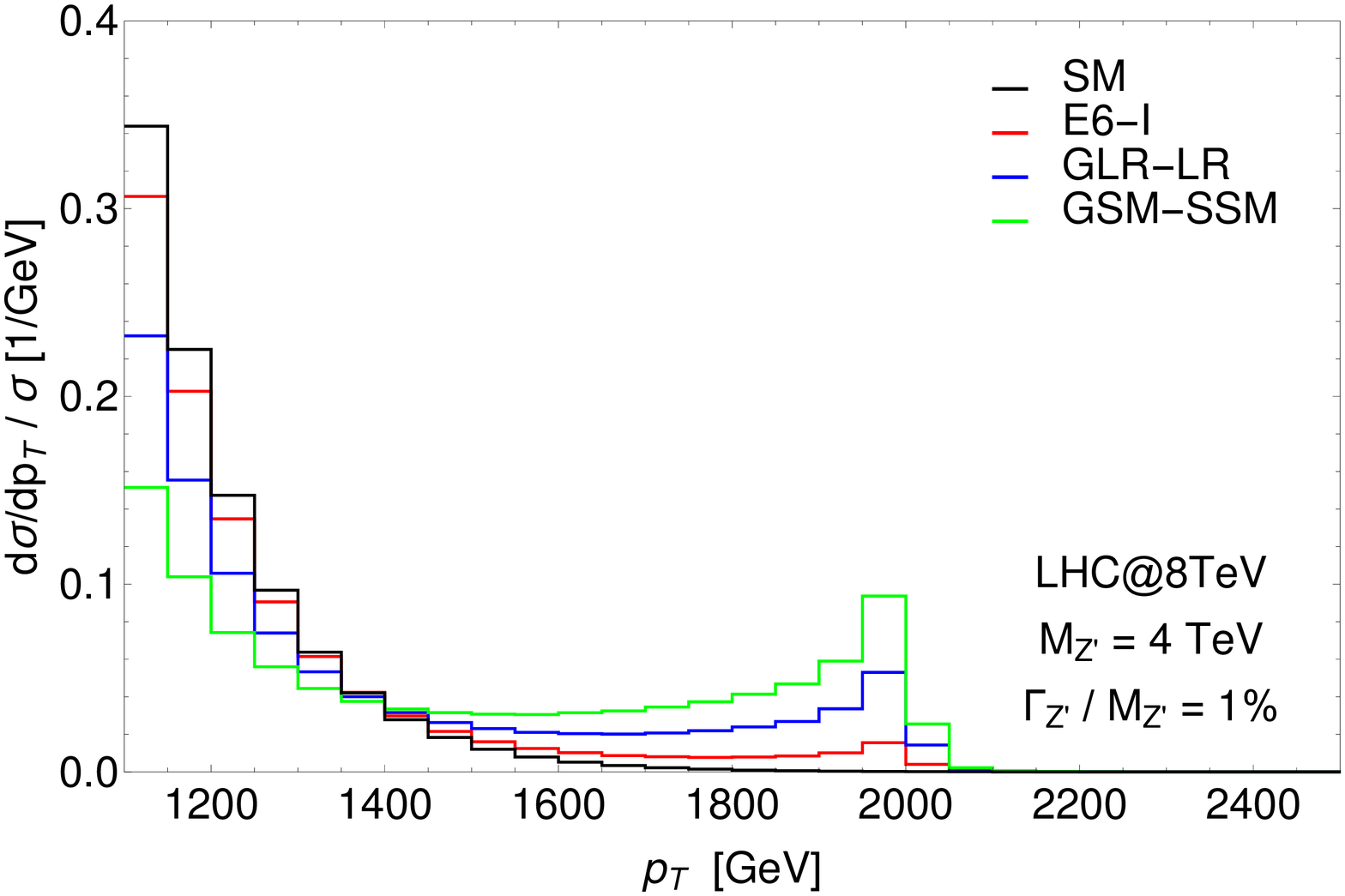}{(b)}
\caption{Normalised distribution in $p_T$ of either lepton as predicted in the SM and in three $Z^\prime$ benchmark models 
with $M_{Z^\prime}=4~{\rm TeV}$ at the 8 TeV LHC.
For all models the width of the resonance has been fixed at 1\% of its mass.
Acceptance cuts are applied ($|\eta|<2.5$), detector efficiencies are not accounted for.
(a) $p_T^{\rm min} = 1000~{\rm GeV}$, (b) $p_T^{\rm min} = 1100~{\rm GeV}$.}
\label{fig:pT_8TeV}
\end{center}
\end{figure}

The observation is found to be is sensitive to the partonic (or collider)
energy. Fig.~\ref{fig:pT_8TeV} (where we have again assumed
$\Gamma_{Z^\prime}/M_{Z^\prime}=1\%$) illustrates that the FP also appears
at 8 TeV for different models and $Z^\prime$ masses considered.

The position of the FP moves with the energy, while maintaining its
feature of model independence.

\subsection{The role of interference}

\begin{figure}[h]
\begin{center}
\includegraphics[width=0.45\textwidth]{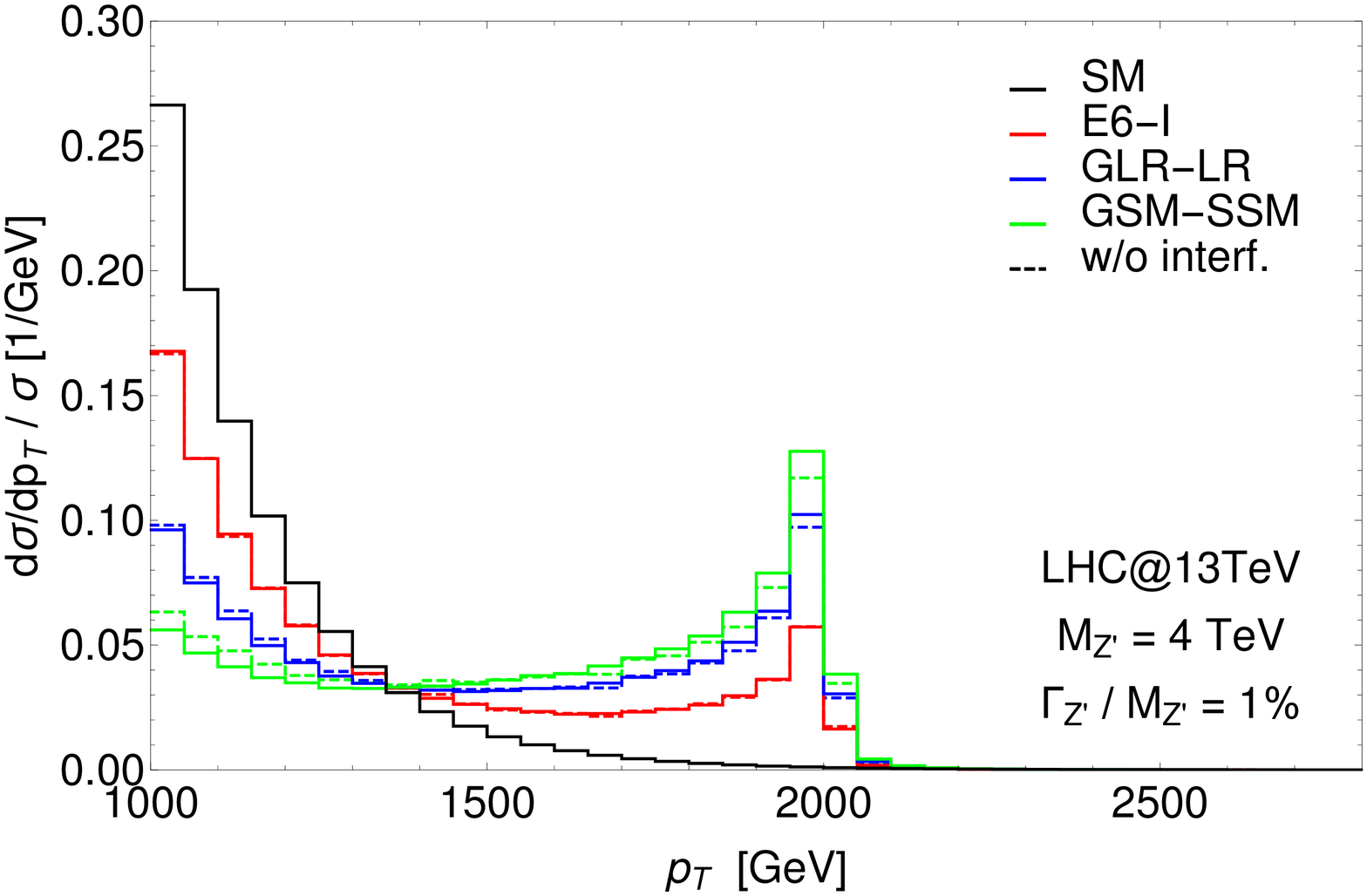}{(a)}
\includegraphics[width=0.45\textwidth]{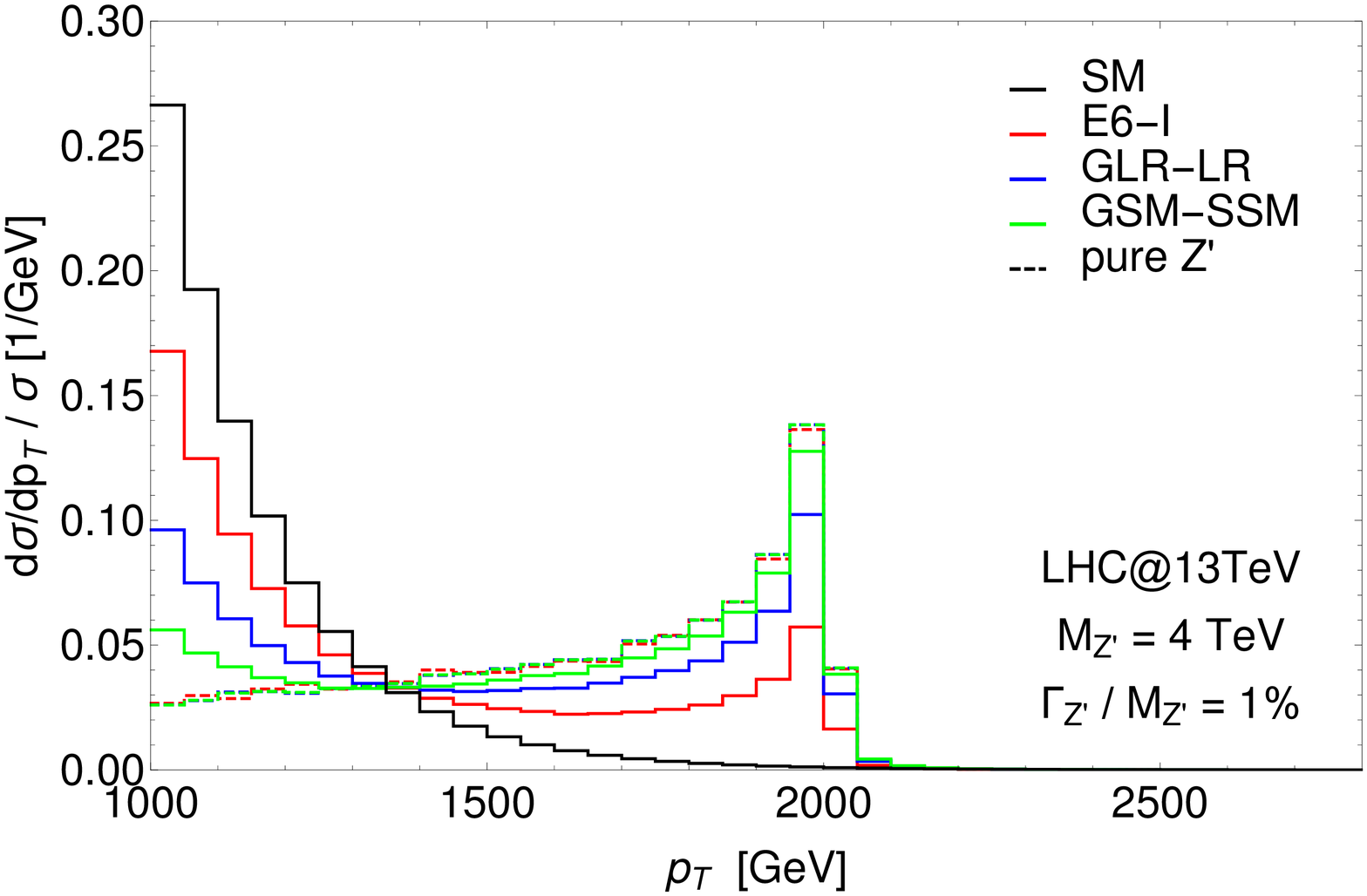}{(b)}
\caption{As in Fig.~\ref{fig:pT_norm}(a) with dashed lines representing (a) the case without the interference terms between the BSM and SM diagrams, and (b) the case of the pure $Z^\prime$ signal.}
\label{fig:pT_interf_norm}
\end{center}
\end{figure}

In this section we explore the role of interference on the observed
FP. In Fig.~\ref{fig:pT_interf_norm}(a), we show the same distribution
as in Fig.~\ref{fig:pT_norm}(a) where, the histograms shown with a
dashed line, correspond to the case where the interference interaction
terms (between the $Z^\prime$ diagram and the $\gamma+Z$ ones) have
been switched off in the MC event generator.
Clearly the contribution of the interference is negligible and it
does not affect the position of the FP.

The same effect is visible in Fig.~\ref{fig:pT_interf_norm}(b) where the
dashed lines represent the $Z^\prime$ signal only, which has been
determined by subtracting the SM background and its interference with the BSM signal. The presence and the
position of the FP are once more unaffected by these changes: all the
curves, representing either the full model or the pure $Z^\prime$
contribution, cross at the same point, demonstrating the stability of
the FP manifestation.
In conclusion, the FP position shows very little dependence on
interference effects, further illustrating the model independent
nature of this result.

\subsection{The role of the width}

\begin{figure}[h]
\begin{center}
\includegraphics[width=0.45\textwidth]{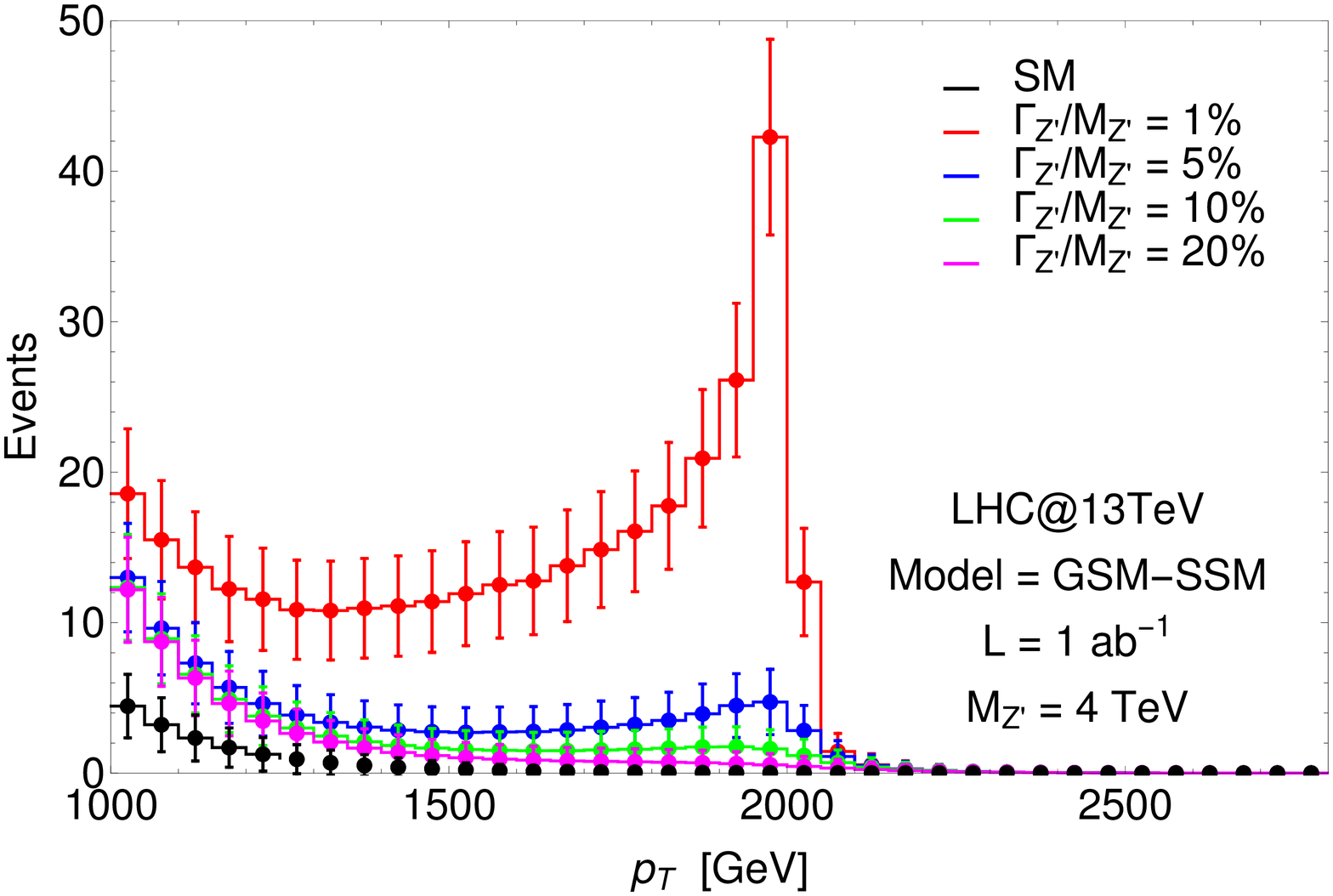}{(a)}
\includegraphics[width=0.45\textwidth]{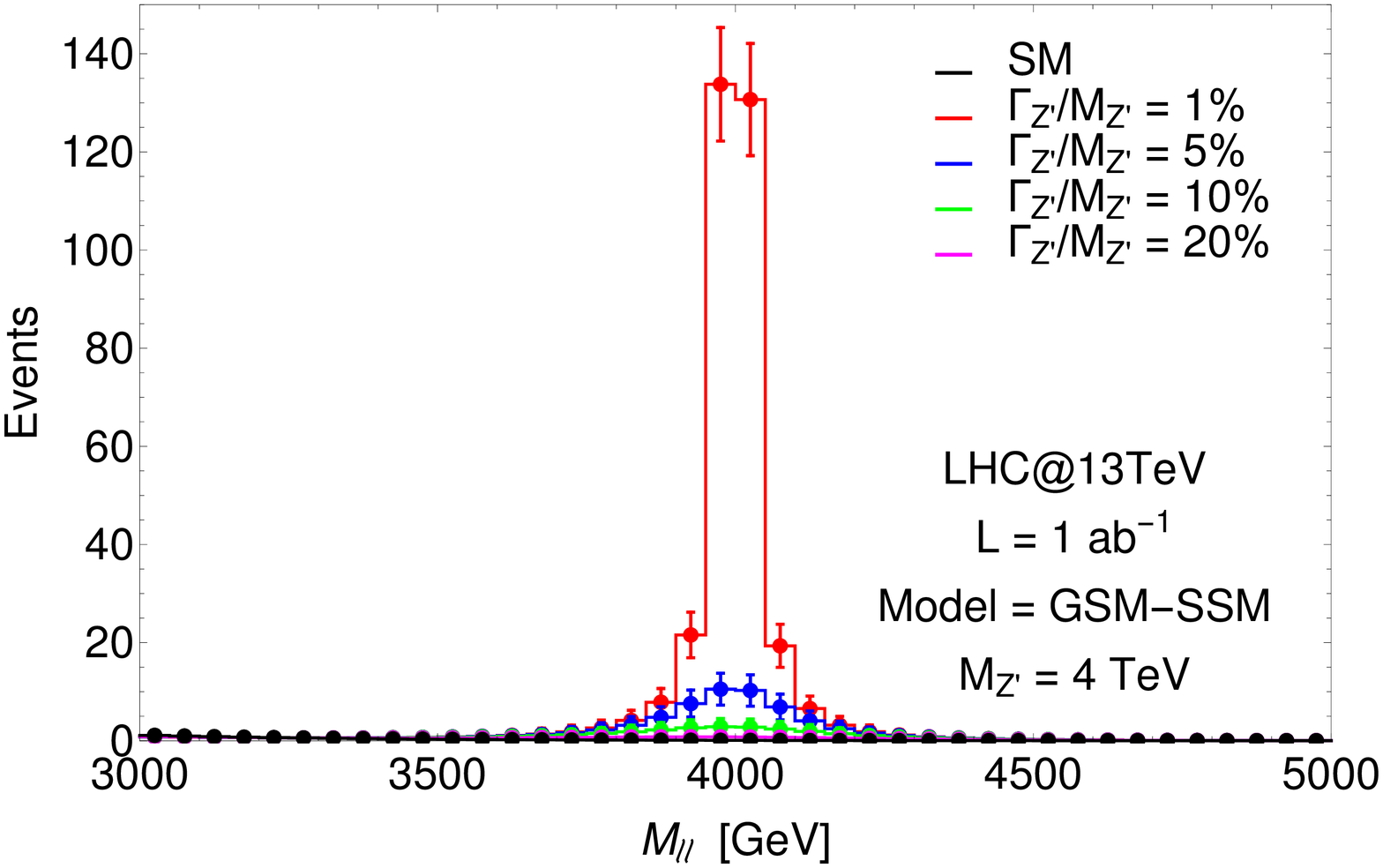}{(b)}
\caption{Distribution of number of events as function of (a) the $p_T$ of either lepton and (b) of the di-lepton invariant mass
as predicted in the SM and in the SSM with $M_{Z^\prime}=4~{\rm TeV}$ at the 13 TeV LHC with $\mathcal{L}=1~ab^{-1}$.
The width of the resonances has been fixed at four different values (1\%, 5\%, 10\% and 20\% of the mass).
Acceptance cuts are applied ($|\eta|<2.5$), no detector efficiencies are accounted for.}
\label{fig:pT_SSM}
\end{center}
\end{figure}

We now consider the affect of varying the width on the FP. For this
purpose, we focus on one specific benchmark, since similar results can
be obtained in the other models. We show in Fig.~\ref{fig:pT_SSM} the
binned distributions of the number of events as function of the lepton
$p_T$ (a) and of the di-lepton system invariant mass (b) for the SSM
model and different choices of the resonance width (1\%, 5\%, 10\% and
20\% of the mass) keeping the mass of the resonance fixed at 4 TeV.
We stress again that in this analysis the width of the
resonance has been enhanced by hand, that is the production cross section
is unchanged, as well as the partial widths into the SM final states.
The branching ratios however scales inversely with the width.
This is representative of a scenario where extra decay channels are 
accessible to the neutral resonance, which is a very common picture 
in many BSM realisations predicting exotic matter.

\begin{figure}[h]
\begin{center}
\includegraphics[width=0.7\textwidth]{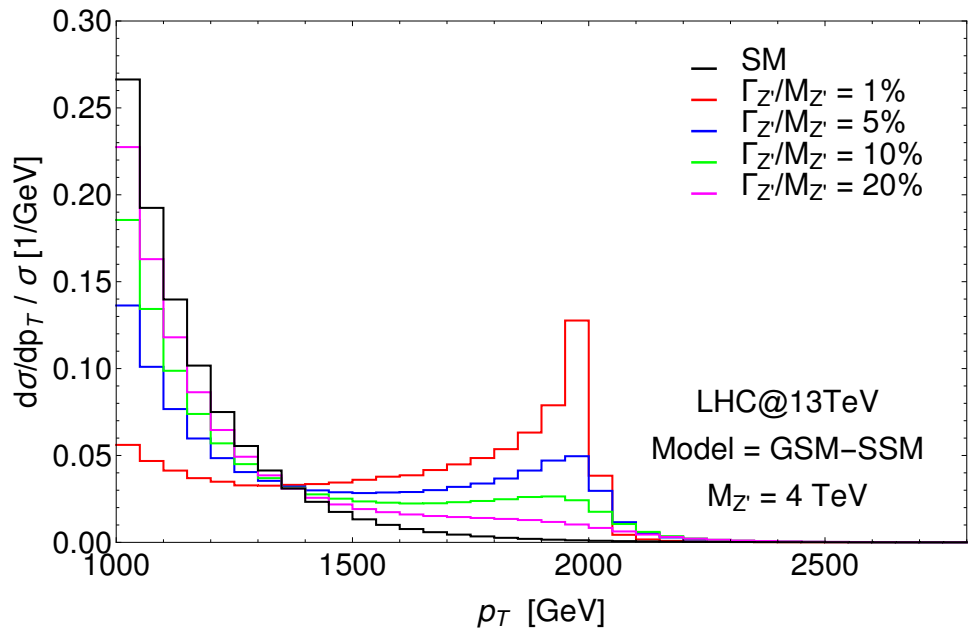}
\caption{Normalized distribution obtained from Fig.~\ref{fig:pT_SSM}(a) with $p_T^{\rm min} = 1000~GeV$.}
\label{fig:pT_SSM_norm}
\end{center}
\end{figure}

In Fig.~\ref{fig:pT_SSM_norm} we illustrate the affect of different
resonance width choices on the FP that appears after the usual
normalisation procedure. The position of the FP can be seen to not
depend on the resonance width. This is the key feature we exploit to
define a new observable that can be used to constrain the resonance
width.

\subsection{The role of the mass}

\begin{figure}[h]
\begin{center}
\includegraphics[width=0.7\textwidth]{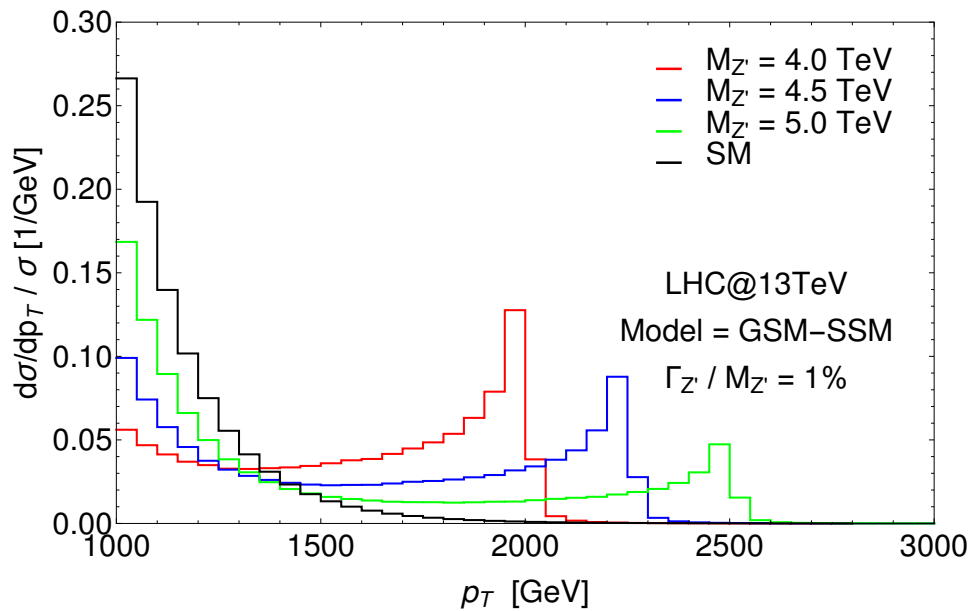}
\caption{Normalised distribution in $p_T$ of either lepton as predicted in the SM and in in the SSM at the 13 TeV LHC.
The mass of the resonances has been fixed at three different values (4.0, 4.5 and 5.0 TeV) while its width has been fixed at 1\% of its mass.
Acceptance cuts are applied ($|\eta|<2.5$), detector efficiencies are not accounted for.}
\label{fig:pT_SSM_mass_norm}
\end{center}
\end{figure}

The effect of varying the $Z^\prime$ resonance mass is shown in the
normalised pt distributions of Fig.~\ref{fig:pT_SSM_mass_norm}. 
The SSM benchmark model is used where we constrain $\Gamma_{Z^\prime}/M_{Z^\prime}=1\%$. 
The position of the FP i.e. the intersection of the model curves with the
SM background, does depend on $Z^\prime$ the mass as expected. 

\subsection{The role of the low $p_T$ cut}

\begin{figure}[h]
\begin{center}
\includegraphics[width=0.45\textwidth]{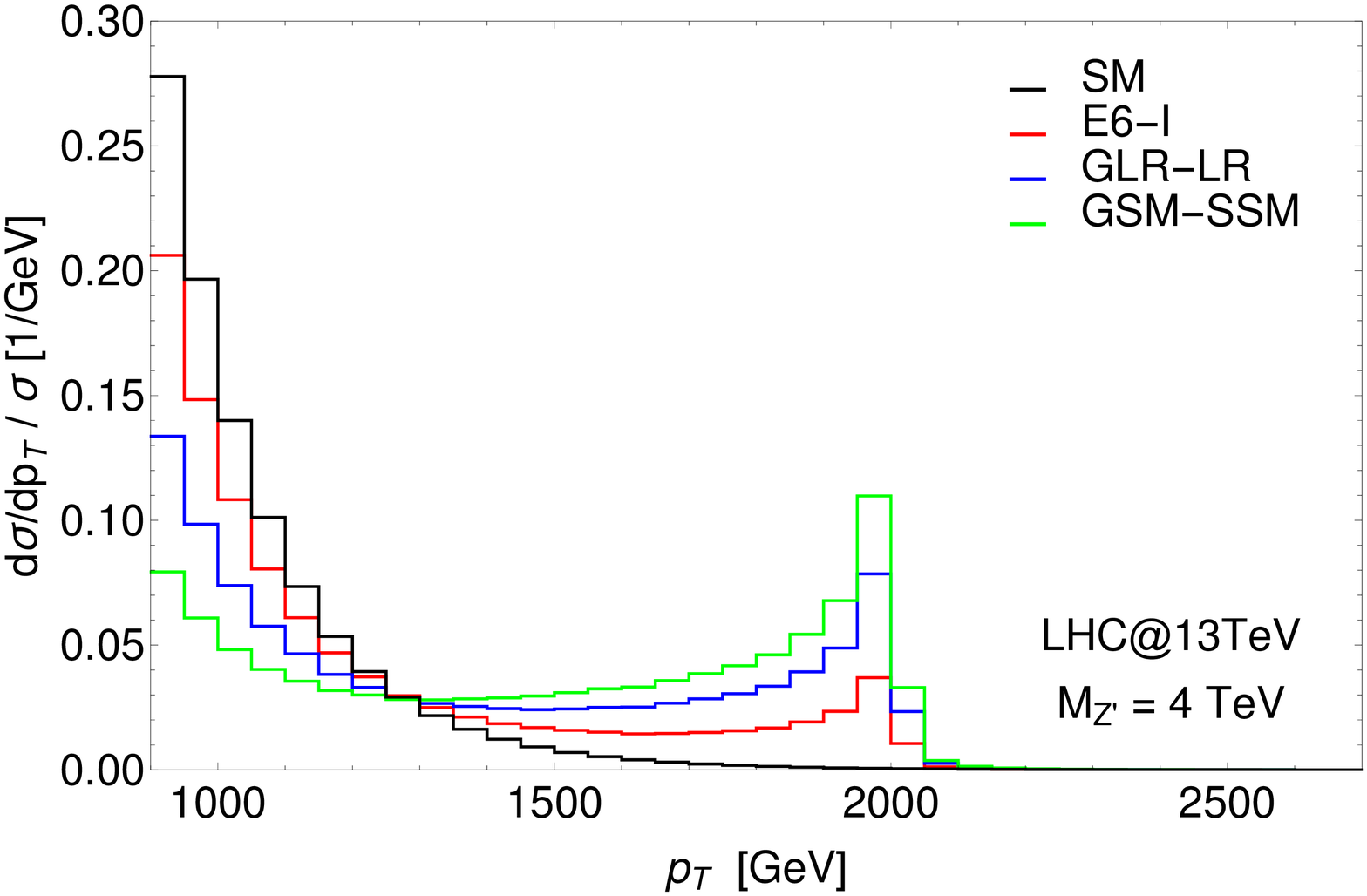}{(a)}
\includegraphics[width=0.45\textwidth]{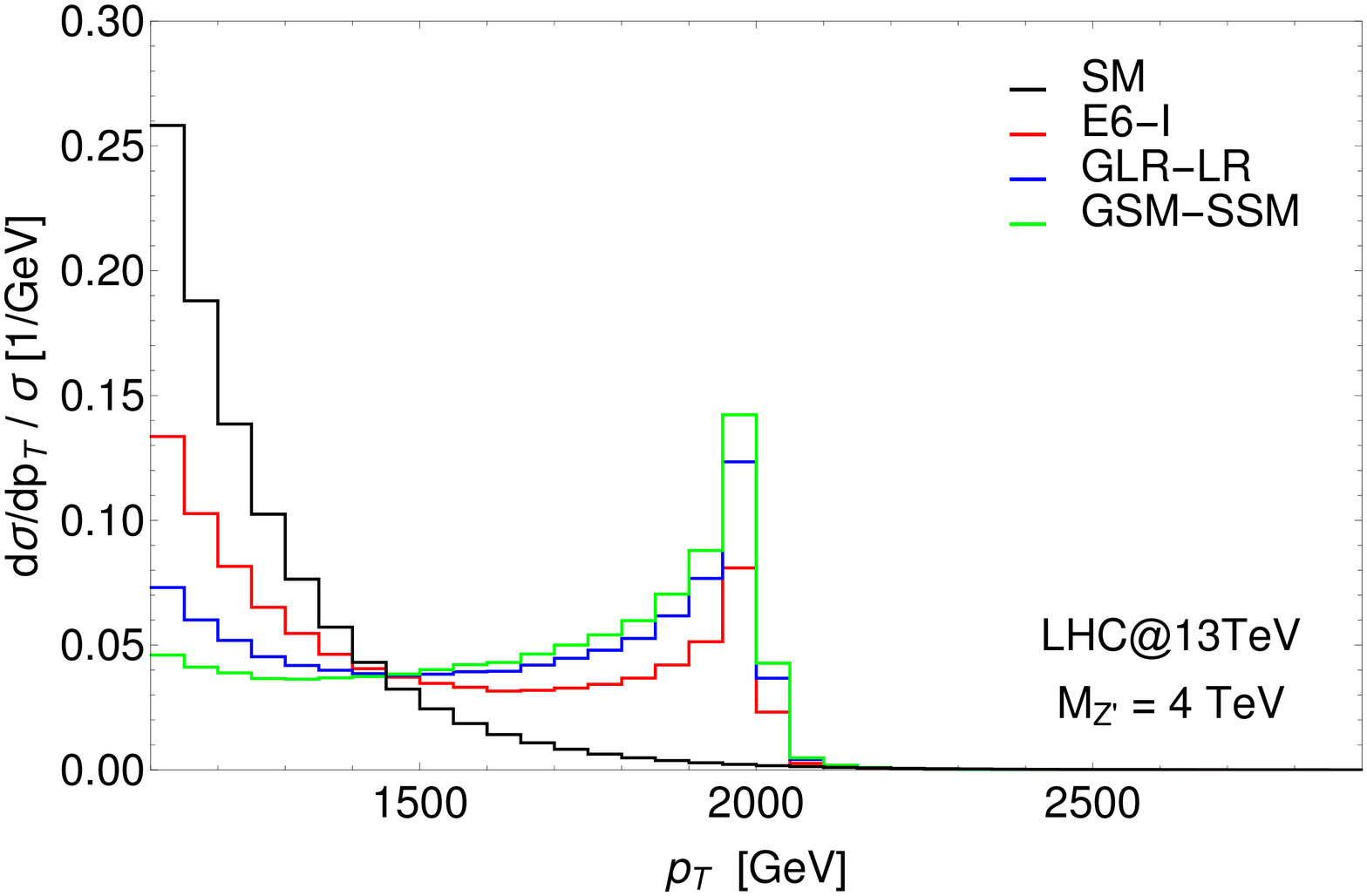}{(b)}
\caption{Normalised distribution in $p_T$ of either lepton as predicted in the SM and in three $Z^\prime$ benchmark models with $M_{Z^\prime}=4~{\rm TeV}$ at the 13 TeV LHC.
For all models the width of the resonance has been fixed at 1\% of its mass.
Acceptance cuts are applied ($|\eta|<2.5$), detector efficiencies are not accounted for.
(a) $p_T^{\rm min} = 900~{\rm GeV}$, (b) $p_T^{\rm min} = 1100~{\rm GeV}$.}
\label{fig:pT_low_cut}
\end{center}
\end{figure}

The main parameter affecting the FP position is, the choice of the low
$p_T$ integration limit, which determines the curves' normalisation
factor. As shown in Fig.~\ref{fig:pT_low_cut} the FP can be seen to
change as a function of which low $p_T$ integration limit is applied.
The two different $p_T$ choices in this figure can also be
compared with the one in Fig.~\ref{fig:pT_norm}(a), where $p_T>1000$ GeV was chosen.

A correlation can be established between the FP location (for a given
$Z^\prime$ mass and LHC energy) and the $p_T^{\rm min}$ cut used for
the normalisation procedure.
{We have observed a numerical relation between the position of the FP
and the choice of  $p_T^{\rm min}$ and have heuristically determined that 
for the LHC at 13 TeV we can assume that the FP position (in GeV) is 
\begin{equation}
 {\rm FP}[{\rm GeV}] \approx p_T^{\rm min} + 10\%\; M_{Z^{\prime}}.
 \label{eq:FP}
\end{equation}
in the accessible range of $Z^\prime$ masses}.
 
\subsection{The role of the $\eta$ cut}
\label{subsec:eta_cut}

\begin{figure}[h]
\begin{center}
\includegraphics[width=0.7\textwidth]{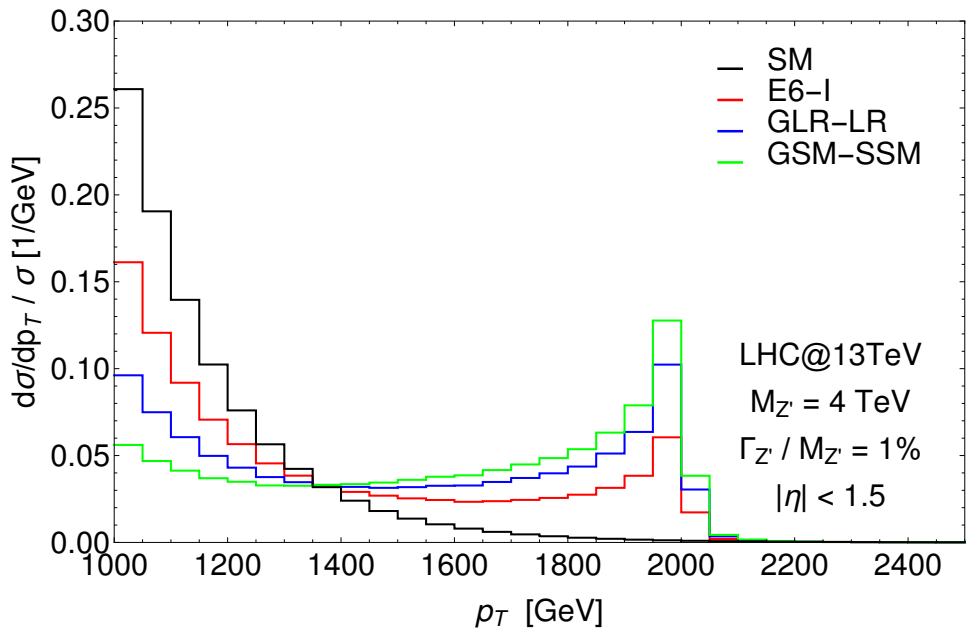}
\caption{Normalised distribution in $p_T$ of either lepton as predicted in the SM and in three $Z^\prime$ benchmark models with $M_{Z^\prime}=4~$TeV.
The width of the resonances has been fixed at 1\% of their mass.
The low $p_T$ cut for our choice of normalisation is $p_T^{\rm min} = 1000~GeV$.
Stronger than default acceptance cuts are applied for these plots ($|\eta|<1.5$), no detector efficiencies are accounted for though. 
Here, $\sqrt s=13$ TeV.}
\label{fig:eta_cut}
\end{center}
\end{figure}

For completeness, in this subsection we show the effect of a change in
selection criterion in the lepton rapidity $\eta^l$. In
Fig.~\ref{fig:eta_cut} we have require($|\eta| < 1.5$) for various
choices of the low $p_T$ cut, to be compared with previous plots. No
observable deviations from previous results are shown and the FP
position does not change.

\section{Constraining $Z^\prime$ widths}
\label{sec:width}

In this section, we will show how the value of the intrinsic
$Z^\prime$ width can be inferred from the use of a novel asymmetry
observable based upon the concept of the FP, as discussed in the
previous sections.

\subsection{Defining a new observable: $A_{\rm FP}$}

For a given collider energy and $Z^\prime$ mass, we have seen that
suitably normalised single-lepton $p_T$ distributions for various
$Z^\prime$ models all have the same magnitude at one point in the
spectrum. We have dubbed this point the Focus Point. The $p_T$ value
associated with it has been shown to not depend upon the intrinsic
$Z^\prime$ width, in any of the models. For a fixed collider
energy and a given $Z^\prime$ mass therefore, it is possible to define a unique
FP that is common to a large class of models.

To define an observable based on the FP feature that can provide
information about the width of the resonance we define two separate
regions in the normalised $p_T$ distribution. The ``Left"
($L$) region going from a fixed $p_T^{\rm min}$ (the low $p_T$ limit
referred to above) up to the FP and the ``Right" ($R$) region going
from the FP up to the last point in the distribution, which we will
assume is $p_T^{\rm max}>M_{Z^\prime}/2$.

We define an asymmetry around the FP, $A_{\rm FP}$, to be the difference between the integrated
normalised distribution in the two regions, divided by the sum of the
two integrations. This can be written 
\begin{equation}
 A_{\rm FP}=\frac{L-R}{L+R}
\end{equation}
with
\begin{equation}
L=\frac{1}{N}\int_{L}\frac{d\sigma}{dp_T}dp_T,\quad\quad
R=\frac{1}{N}\int_{R}\frac{d\sigma}{dp_T}dp_T,
\end{equation}
where the two domains $L$ and $R$ are chosen as described above, i.e.,
$L=\left[p_T^{\rm min},{\rm FP}\right]$, $R=\left[{\rm FP},p_T^{\rm max}\right]$, 
with FP the FP position in the $p_T$ axis, and $N$
the total number of events in the $(L+R)$ region that we have also
used for the normalization procedure. The expression we have derived
for the new observable is notionally very similar to the
Forward-Backward Asymmetry ($A_{\rm FB}$)~\cite{Accomando:2015cfa, Accomando:2015pqa, Accomando:2015ava}.
In this sense, the formula for the statistical error
on the $A_{\rm FP}$ observable is analogous to the one for the $A_{\rm FB}$, thus:

\begin{equation}
 \Delta A_{\rm FP}=\sqrt{\frac{1-A_{\rm FP}^2}{N}},
\end{equation}

This $A_{\rm FP}$ observable can be used to estimate the width of the
$Z^\prime$ resonance, with the positive feature of being unbiased by
systematics and assumptions intrinsic to shape dependent fitting procedures (such as
assuming a Breit-Wigner resonance structure in the the di-lepton
invariant mass spectrum)
Thus, we are going to estimate the $A_{\rm FP}$ values for different
$Z^\prime$ model and width choices, at the 13 TeV LHC for
various $Z^\prime$ masses. At this point, it is important to mention
that the definition of the $L$ and $R$ regions is crucial for a correct
analysis of the results. The precise steps to follow are: (i)
 extraction of the mass of the resonance from the di-lepton invariant
 mass, possibly combined with the location of the maximum of the
 $p_T$ distribution (which roughly coincides with $M_{Z^\prime}/2$); 
 (ii) definition of the FP position according to Eq.~(\ref{eq:FP}).

While $p_T^{\rm max}$ is essentially defined to be any point in transverse momentum past $M_{Z^\prime}/2$ 
(as seen in the various distributions that we have presented, the drop beyond this point is dramatic), we have some freedom in the choice of 
$p_T^{\rm min}$. For example, a high $p_T^{\rm min}$ would maximise the sensitivity to any BSM physics while a low $p_T^{\rm min}$ would maximise 
the sensitivity to different BSM scenarios. As discovery of some BSM physics is assumed to have already occurred from analysis of the 
$M_{ll}$ spectrum, for our purposes, a low $p_T^{\rm min}$ is indeed more appropriate. 

In Tabs.~\ref{tab:table1}--\ref{tab:table2} we show the calculated $A_{\rm FP}$ observable for the SM background and for the usual benchmark models assuming different widths. 
We consider two values for the $Z^\prime$ mass ($M_{Z^\prime}=4~{\rm TeV}$ and $M_{Z^\prime}=5~{\rm TeV}$) 
and three possible choices for the $p_T^{\rm min}$ for each mass. In general, as expected, as we move up the $p_T^{\rm min}$ (and consequently the FP location) we have more sensitivity to the presence of BSM
physics while going in the opposite direction leads to an enhancement of the sensitivity to the $Z^\prime$ boson width.

The statistical errors are also reported in the two tables and they are obtained for an integrated luminosity of 1 and 3 ab$^{-1}$ respectively.
The statistical error represents the dominant uncertainty in the $A_{\rm FP}$ observable.
Being a ratio of cross sections systematic uncertainties are indeed expected to cancel partially. 
We give two examples of the expected size of the PDF uncertainty, to compare with the central value and the statistical error taken from Tab.~\ref{tab:table2}.

\begin{align*}
  &M_{Z^\prime} = 5~{\rm TeV},~\Gamma_{Z^\prime}/M_{Z^\prime} = 5\%,~p_T^{\rm min} = 1.2~{\rm TeV},\\ 
  &(A_{\rm FP} \pm \Delta_{\rm stat} \pm \Delta_{\rm PDF})_{\rm SM} = 0.87 \pm 0.07 \pm 0.01,\\
  &(A_{\rm FP} \pm \Delta_{\rm stat} \pm \Delta_{\rm PDF})_{\rm SSM} = 0.44 \pm 0.12 \pm 0.06.
  \label{eq:PDF_error}
\end{align*}

\begin{table}[h]
\begin{center}
\begin{tabular}{|c||c|c|c|c|}
 \hline
 \multicolumn{5}{|c|}{$M_{Z^\prime}=4~{\rm TeV}$} \\
 \hline
 Model & $\Gamma_{Z^\prime} / M_{Z^\prime} = 1\%$ & $\Gamma_{Z^\prime} / M_{Z^\prime} = 5\%$ & $\Gamma_{Z^\prime} / M_{Z^\prime} = 10\%$ & $\Gamma_{Z^\prime} / M_{Z^\prime} = 20\%$\\
 \Xhline{4\arrayrulewidth}
 \multicolumn{5}{|c|}{$p_T^{\rm min}=900~{\rm GeV}$} \\
 \hline
 SM & \multicolumn{4}{c|}{0.82$\pm$0.05} \\
 \hline
 $E_6^I$ & 0.44$\pm$0.07 & 0.72$\pm$0.06 & 0.77$\pm$0.06 & 0.80$\pm$0.06 \\
 LR & 0.02$\pm$0.07 & 0.55$\pm$0.07 & 0.68$\pm$0.07 & 0.76$\pm$0.06 \\
 SSM & -0.29$\pm$0.05 & 0.26$\pm$0.08 & 0.50$\pm$0.08 & 0.67$\pm$0.07 \\
 \hline
 \multicolumn{5}{|c|}{ $p_T^{\rm min}=1000~{\rm GeV}$} \\
 \hline
 SM & \multicolumn{4}{c|}{0.81$\pm$0.08} \\
 \hline
 $E_6^I$ & 0.27$\pm$0.10 & 0.65$\pm$0.09 & 0.72$\pm$0.09 & 0.77$\pm$0.08 \\
 LR & -0.14$\pm$0.07 & 0.40$\pm$0.10 & 0.58$\pm$0.10 & 0.70$\pm$0.09 \\
 SSM & -0.37$\pm$0.05 & 0.06$\pm$0.10 & 0.33$\pm$0.12 & 0.56$\pm$0.11 \\
 \hline
 \multicolumn{5}{|c|}{$p_T^{\rm min}=1100~{\rm GeV}$} \\
 \hline
 SM & \multicolumn{4}{c|}{0.79$\pm$0.11} \\
 \hline
 $E_6^I$ & 0.12$\pm$0.12 & 0.57$\pm$0.13 & 0.68$\pm$0.12 & 0.74$\pm$0.12 \\
 LR & -0.22$\pm$0.08 & 0.25$\pm$0.14 & 0.47$\pm$0.14 & 0.64$\pm$0.13 \\
 SSM & -0.38$\pm$0.05 & -0.08$\pm$0.12 & 0.16$\pm$0.15 & 0.43$\pm$0.16 \\
 \hline
\end{tabular}
\end{center}
\caption{$A_{\rm FP}$ and its statistical error for the SM and three benchmark models with $M_{Z^\prime}=4~{\rm TeV}$ and four different widths 
repeated for three choices of $p_T^{\rm min}$, for the LHC at 13 TeV and $\mathcal{L}=1~{\rm ab}^{-1}$. 
The FP position is obtained following Eq.~\ref{eq:FP}.}
\label{tab:table1}
\end{table}

\begin{table}[h]
\begin{center}
\begin{tabular}{|c||c|c|c|c|}
 \hline
 \multicolumn{5}{|c|}{$M_{Z^\prime}=5~{\rm TeV}$} \\
 \hline
 Model & $\Gamma_{Z^\prime} / M_{Z^\prime} = 1\%$ & $\Gamma_{Z^\prime} / M_{Z^\prime} = 5\%$ & $\Gamma_{Z^\prime} / M_{Z^\prime} = 10\%$ & $\Gamma_{Z^\prime} / M_{Z^\prime} = 20\%$\\
 \Xhline{4\arrayrulewidth}
 \multicolumn{5}{|c|}{$p_T^{\rm min}=1100~{\rm GeV}$} \\
 \hline
 SM & \multicolumn{4}{c|}{0.88$\pm$0.05} \\
 \hline
 $E_6^I$ & 0.71$\pm$0.07 & 0.84$\pm$0.06 & 0.85$\pm$0.05 & 0.87$\pm$0.05 \\
 LR & 0.40$\pm$0.08 & 0.76$\pm$0.07 & 0.82$\pm$0.06 & 0.85$\pm$0.06 \\
 SSM & 0.04$\pm$0.08 & 0.60$\pm$0.08 & 0.74$\pm$0.07 & 0.82$\pm$0.06 \\
 \hline
 \multicolumn{5}{|c|}{ $p_T^{\rm min}=1200~{\rm GeV}$} \\
 \hline
 SM & \multicolumn{4}{c|}{0.87$\pm$0.07} \\
 \hline
 $E_6^I$ & 0.62$\pm$0.10 & 0.81$\pm$0.08 & 0.84$\pm$0.07 & 0.85$\pm$0.07 \\
 LR & 0.22$\pm$0.10 & 0.68$\pm$ 0.10 & 0.77$\pm$0.09 & 0.83$\pm$0.08 \\
 SSM & -0.14$\pm$0.09 & 0.44$\pm$0.12 & 0.64$\pm$0.11 & 0.77$\pm$0.10 \\
 \hline
 \multicolumn{5}{|c|}{$p_T^{\rm min}=1300~{\rm GeV}$} \\
 \hline
 SM & \multicolumn{4}{c|}{0.86$\pm$0.09} \\
 \hline
 $E_6^I$ & 0.50$\pm$0.14 & 0.77$\pm$0.11 & 0.81$\pm$0.10 & 0.84$\pm$0.10 \\
 LR & 0.06$\pm$0.12 & 0.58$\pm$0.14 & 0.72$\pm$0.13 & 0.80$\pm$0.11 \\
 SSM & -0.24$\pm$0.09 & 0.27$\pm$0.16 & 0.52$\pm$0.16 & 0.70$\pm$0.14 \\
 \hline
\end{tabular}
\end{center}
\caption{$A_{\rm FP}$ and its statistical error for the SM and three benchmark models with $M_{Z^\prime}=5~{\rm TeV}$ and four different widths 
repeated for three choices of $p_T^{\rm min}$, for the LHC at 13 TeV and $\mathcal{L}=3~{\rm ab}^{-1}$. 
The FP position is obtained following Eq.~\ref{eq:FP}.}
\label{tab:table2}
\end{table}

\subsection{Sensitivity of the $A_{\rm FP}$ observable}
\label{subsec:sensitivity}

\begin{figure}[h]
\begin{center}
\includegraphics[width=0.45\textwidth]{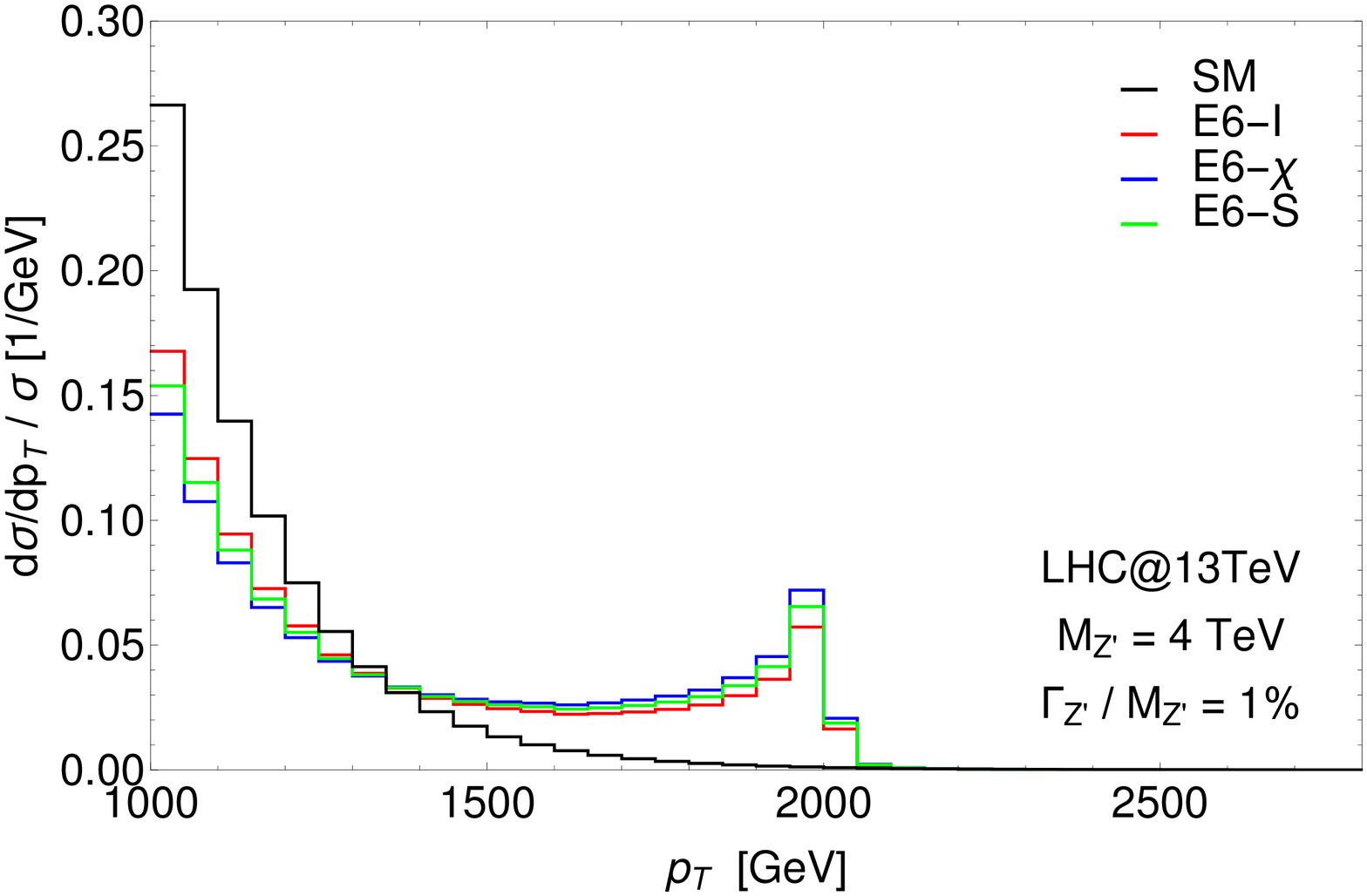}{(a)}
\includegraphics[width=0.45\textwidth]{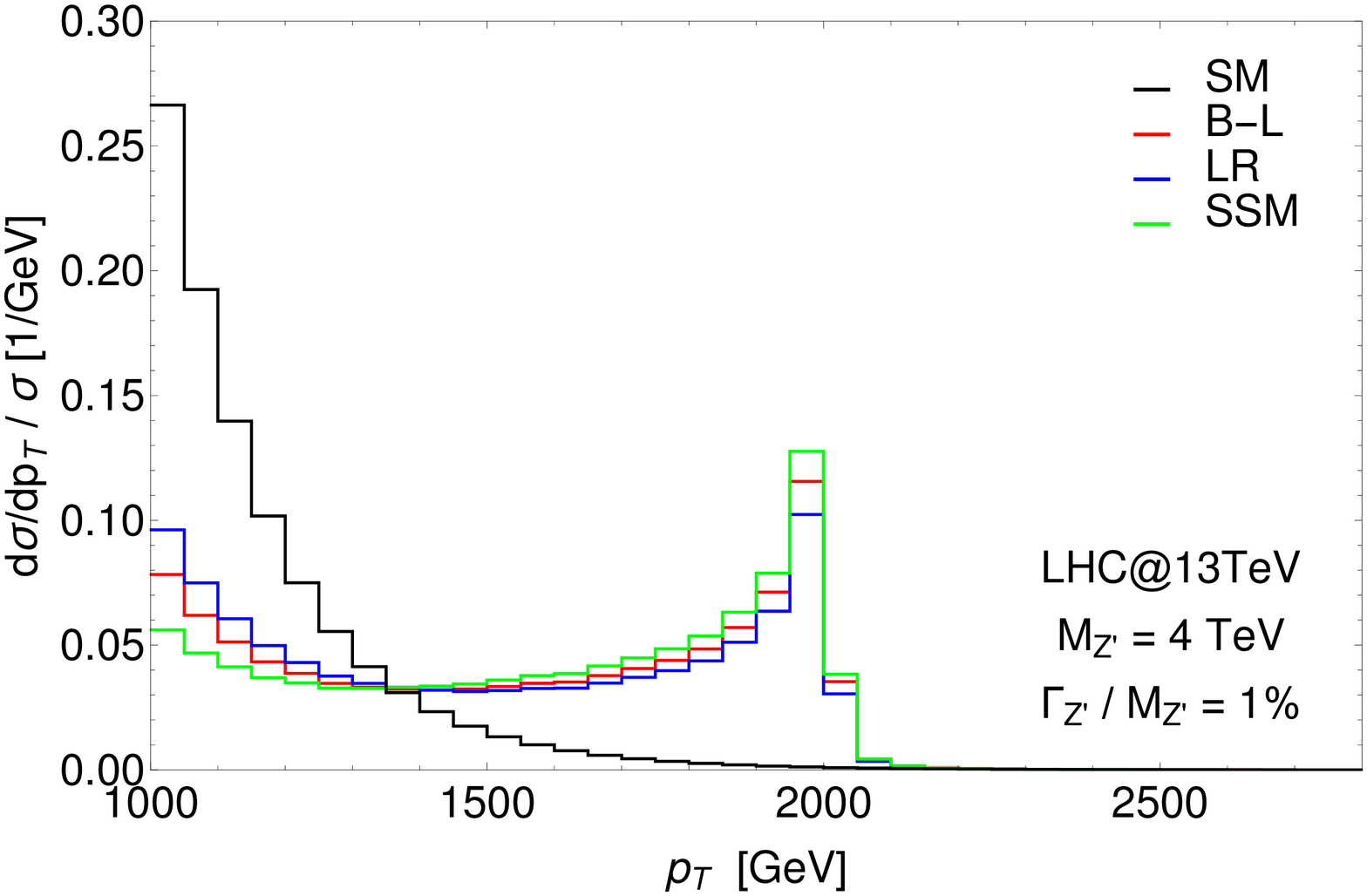}{(b)}
\caption{Normalised distribution in $p_T$ of either lepton as predicted in the SM (black) and in three $Z^\prime$ benchmark models (coloured) within 
the $E_6$ class (a) and GLR and GSM classes (b) with $M_{Z^\prime}=4~$TeV and $p_T^{\rm min} = 1000~GeV$.
The width of the resonances has been fixed at 1\% of their mass.
Acceptance cuts are applied ($|\eta|<2.5$), no detector efficiencies are accounted for. Here, $\sqrt s=13$ TeV.}
\label{fig:pT_E6_LR}
\end{center}
\end{figure}

In this section we want to explore in more detail the potential of the new $A_{\rm FP}$ observable in discriminating amongst
different $Z^\prime$ models. We begin by comparing BSM scenarios within the same class. We do so in Fig.~\ref{fig:pT_E6_LR}, 
where we show the usual normalised $p_T$ distribution.

\begin{figure}[h]
\begin{center}
\includegraphics[width=0.45\textwidth]{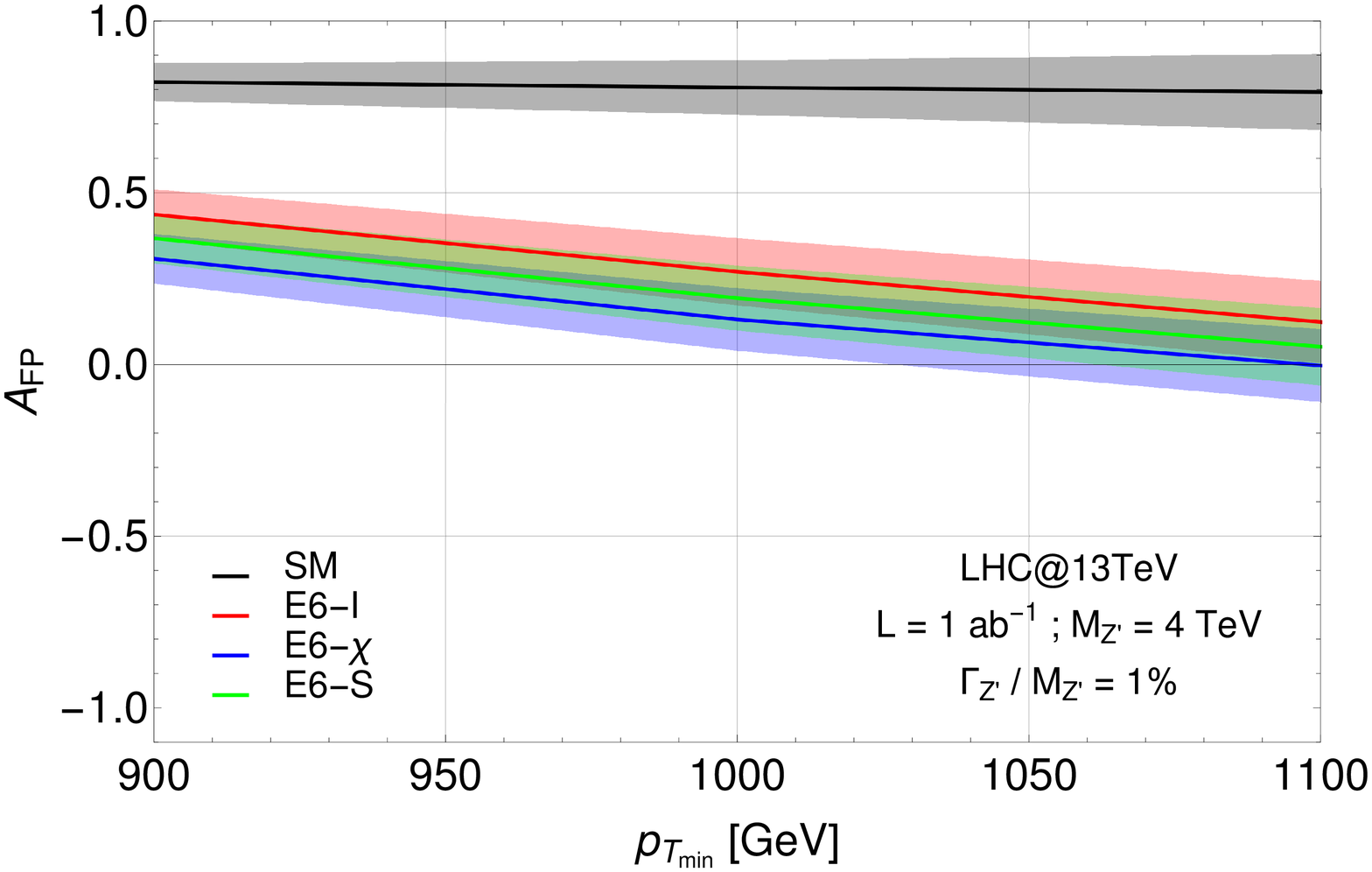}{(a)}
\includegraphics[width=0.45\textwidth]{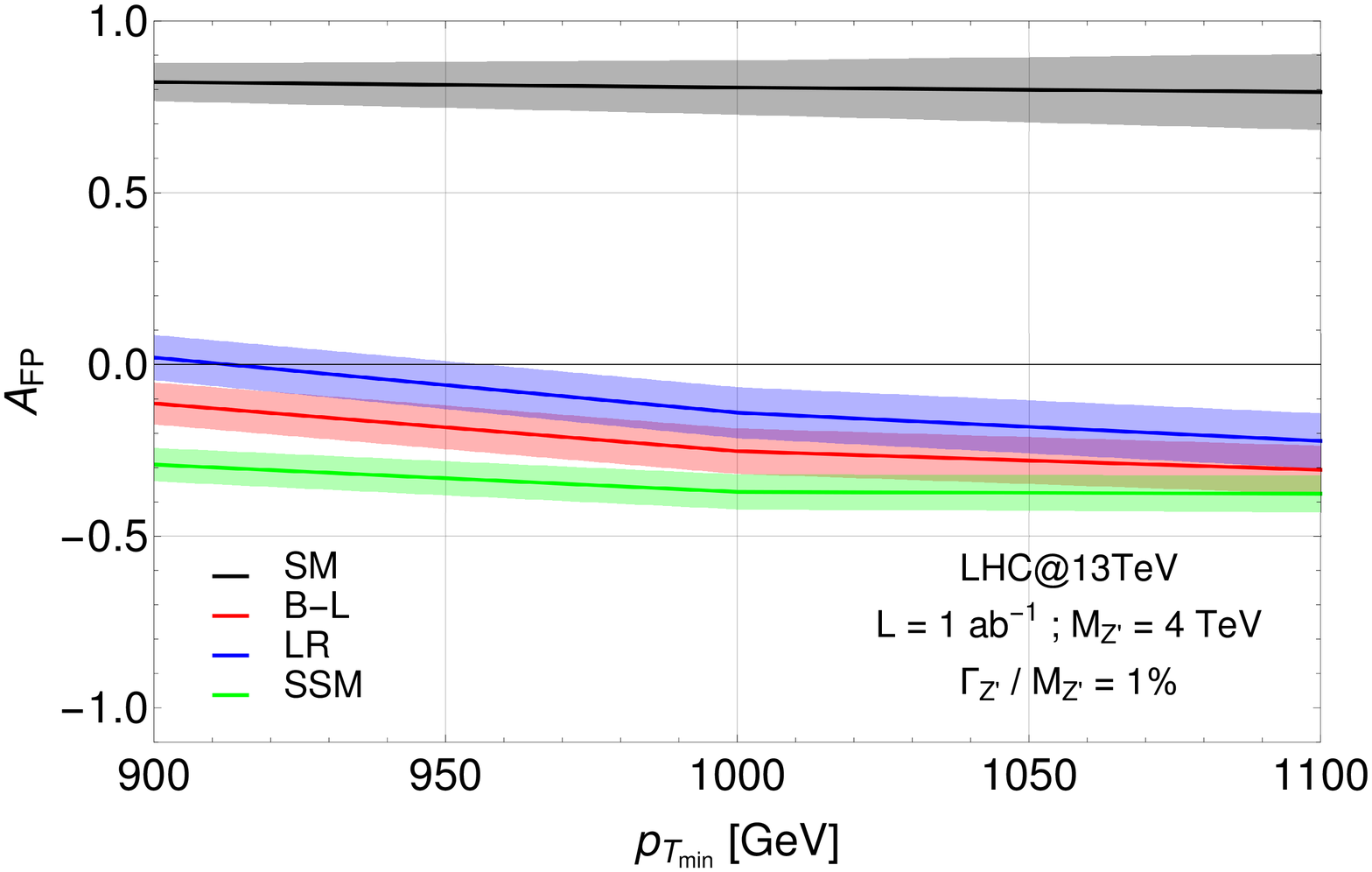}{(b)}
\caption{$A_{\rm FP}$ central value and statistical 1$\sigma$ error band as function of $p_T^{\rm min}$ for the LHC at 13 TeV and 
$\mathcal{L}=1~{\rm ab}^{-1}$.
The black line represents the SM while the coloured lines represent three benchmark in the $E_6$ class (a) and GLR and GSM classes (b). 
The mass of the $Z^\prime$ boson is fixed at 4 TeV and its width has been fixed $\Gamma / M = 1\%$.
The values for the FPs are chosen in accordance to the tables above.}
\label{fig:AFP_models}
\end{center}
\end{figure}

The distributions of the models in the $E_6$ class present clear similarities and the same behaviour is shown in the models belonging to the $LR$ class. 
In Fig.~\ref{fig:AFP_models}, we are showing the $A_{\rm FP}$ and its statistical error as function of the $p_T^{\rm min}$ cut.

For what we can see, $Z^\prime$ models in the same class have similar values for $A_{\rm FP}$, all falling within the error bars already 
for $Z^\prime$ masses of 4 TeV and narrow resonances. This is definitely true for benchmarks in the $E_6$ class and a similar behaviour is shown 
for two GLR benchmarks as well ($LR$ and $B-L$). However, as the resonance mass or width increases, the differences between models tend to disappear.
This, in essence, suggests that we cannot use this observable to discriminate between models within the same class.

\begin{figure}[h]
\begin{center}
\includegraphics[width=0.45\textwidth]{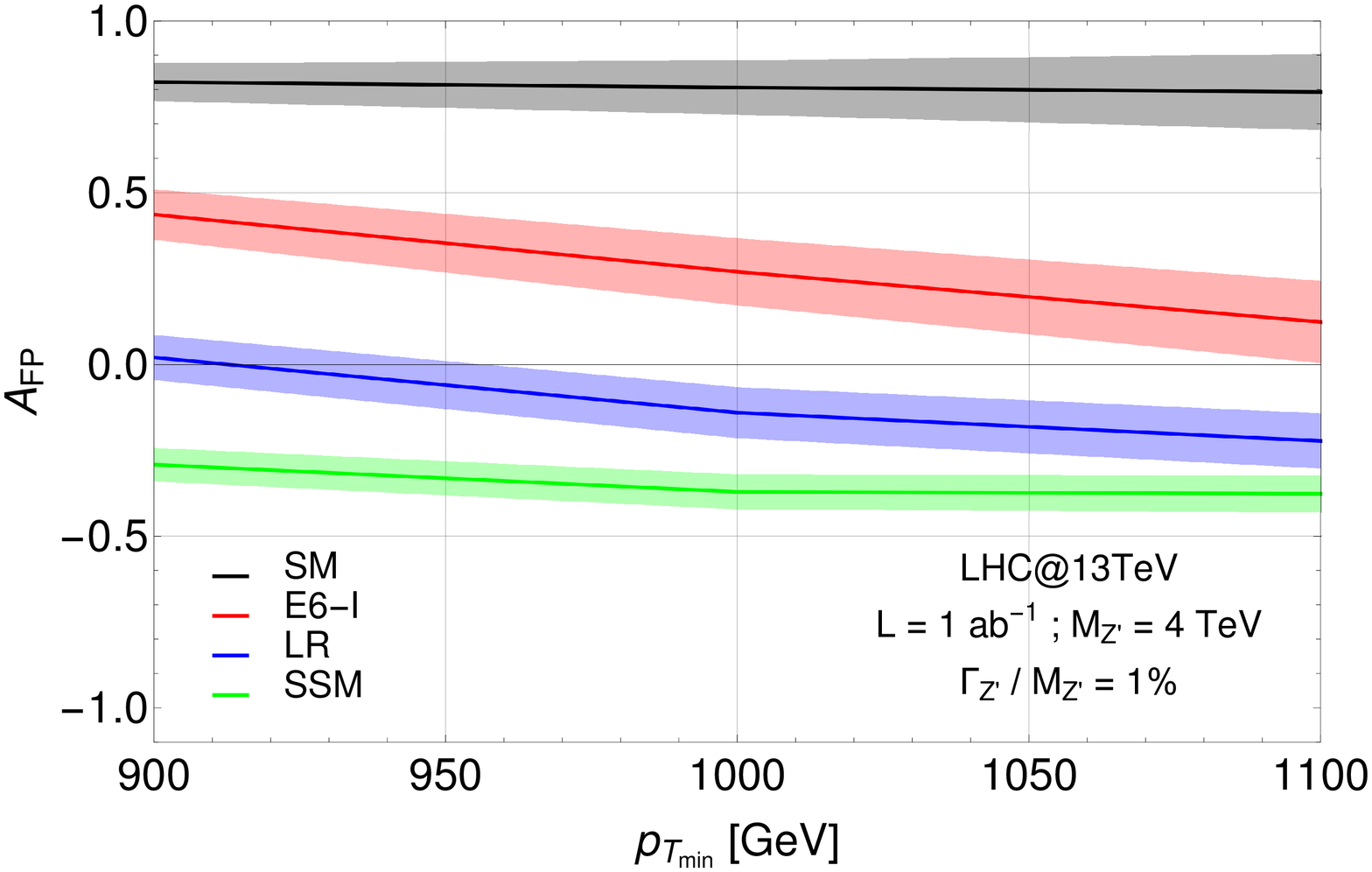}{(a)}
\includegraphics[width=0.45\textwidth]{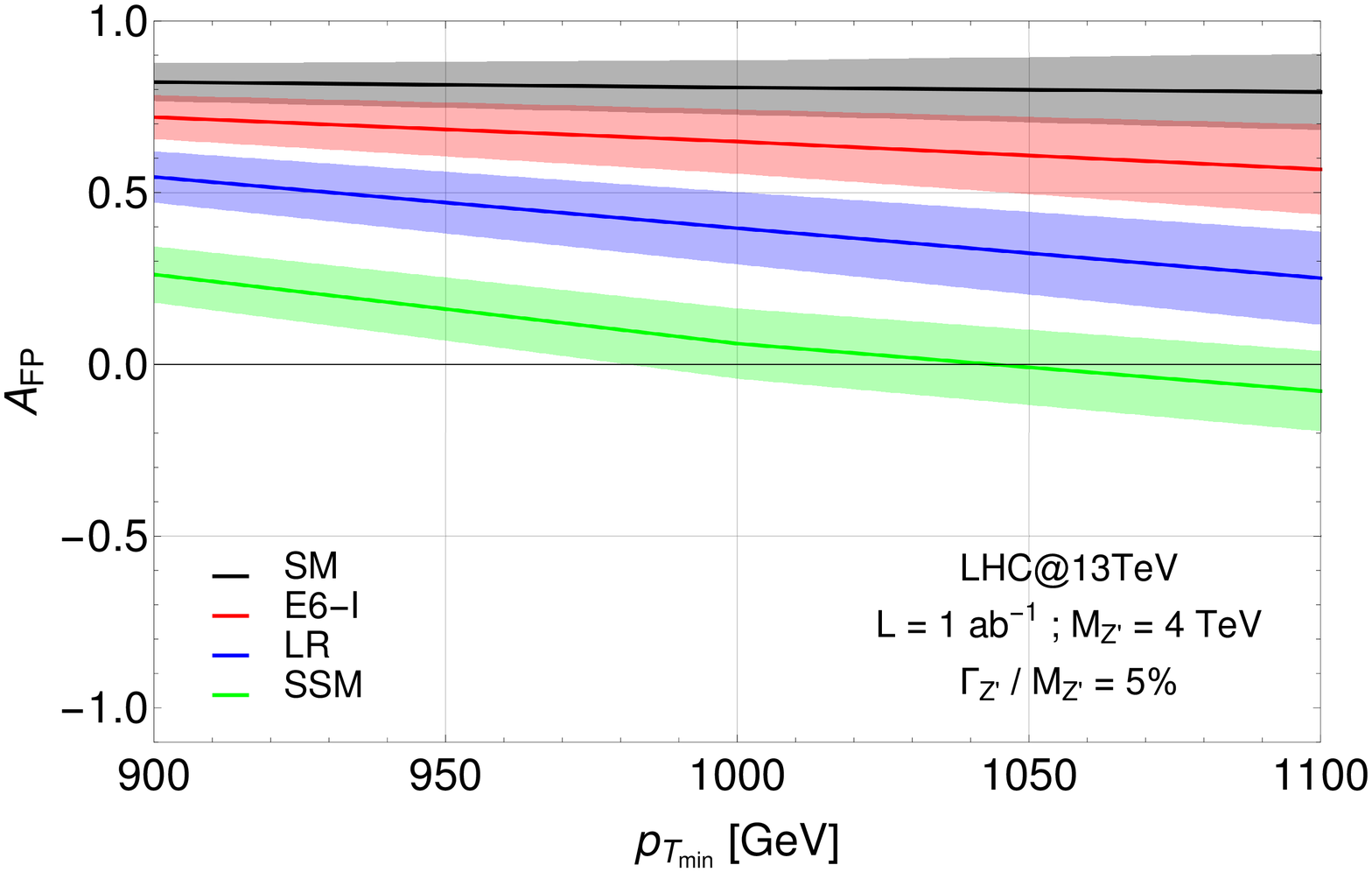}{(b)}
\includegraphics[width=0.45\textwidth]{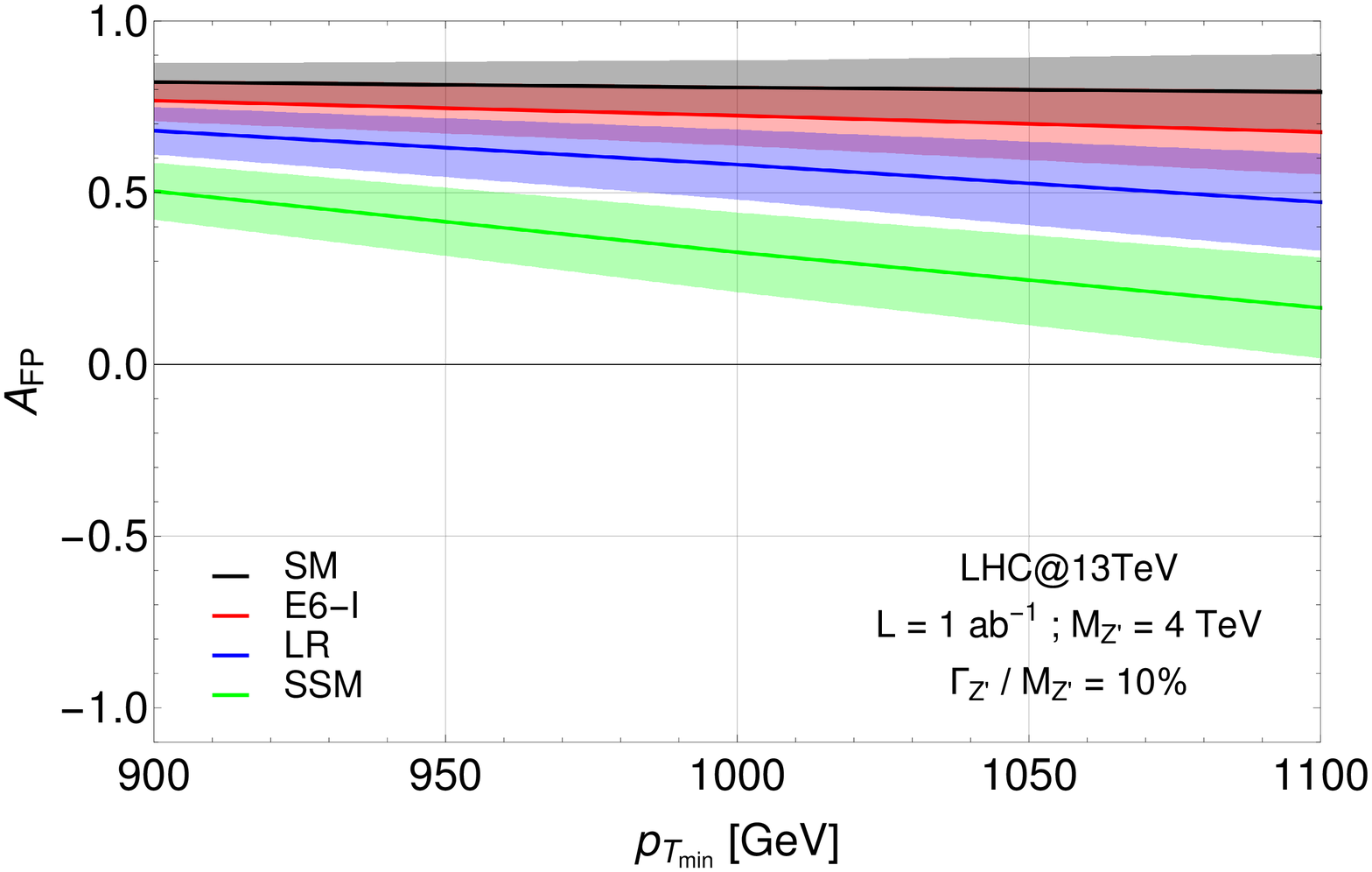}{(c)}
\includegraphics[width=0.45\textwidth]{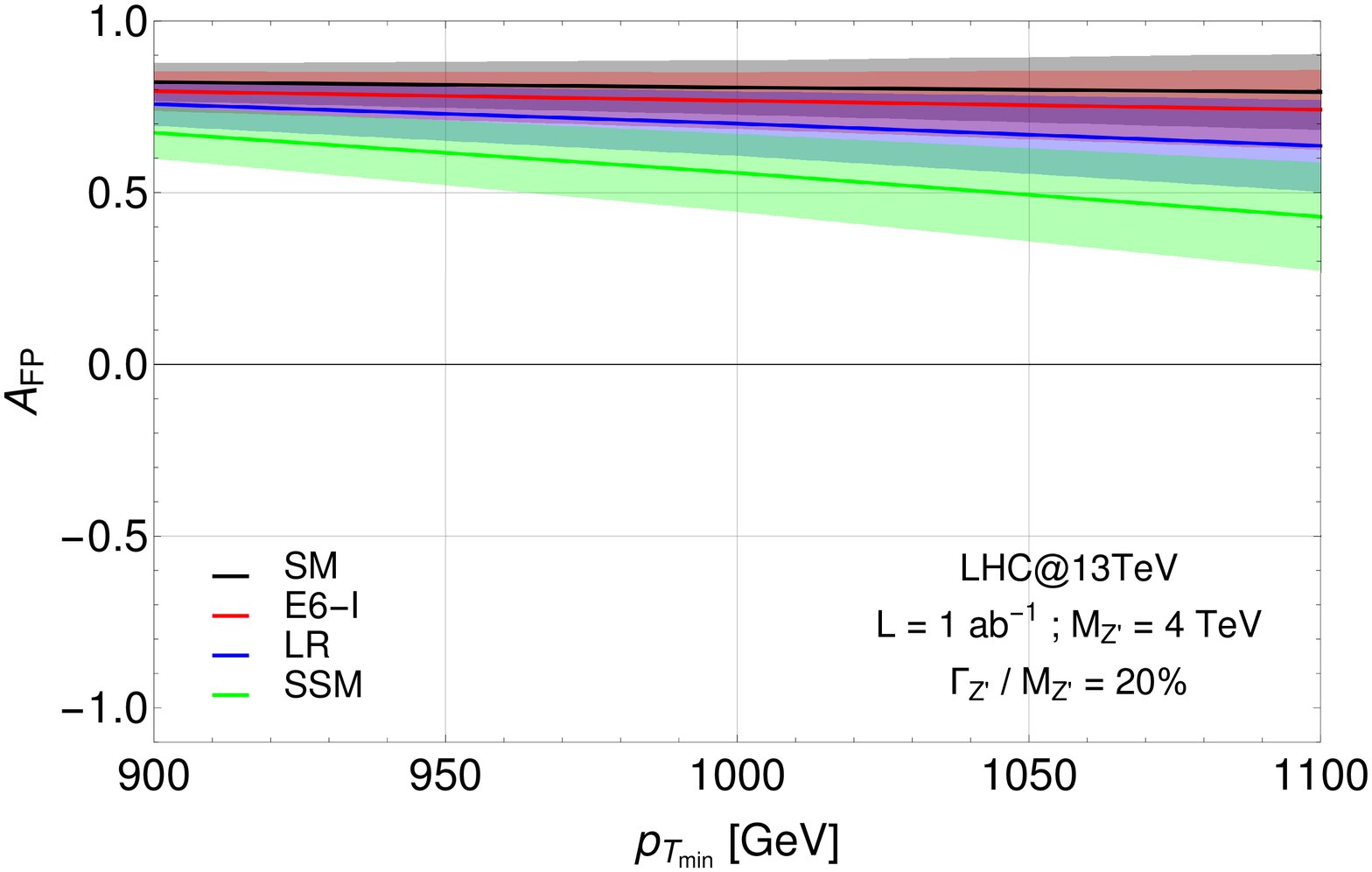}{(d)}
\caption{$A_{\rm FP}$ central value and statistical 1$\sigma$ error band as function of $p_T^{\rm min}$ cut for the LHC at 13 TeV 
and $\mathcal{L}=1~{\rm fb}^{-1}$. The black line represents the SM while the coloured lines represent the three benchmark models.
The mass of the $Z^\prime$ boson is fixed at 4 TeV while its width over mass ratio $\Gamma_{Z^\prime} / M_{Z^\prime}$ has been fixed 
to 1\% (a), 5\% (b), 10\% (c) and 20\% (d). The values for the FP are chosen in accordance to the tables above.}
\label{fig:AFP_models_4}
\end{center}
\end{figure}

\begin{figure}[h]
\begin{center}
\includegraphics[width=0.45\textwidth]{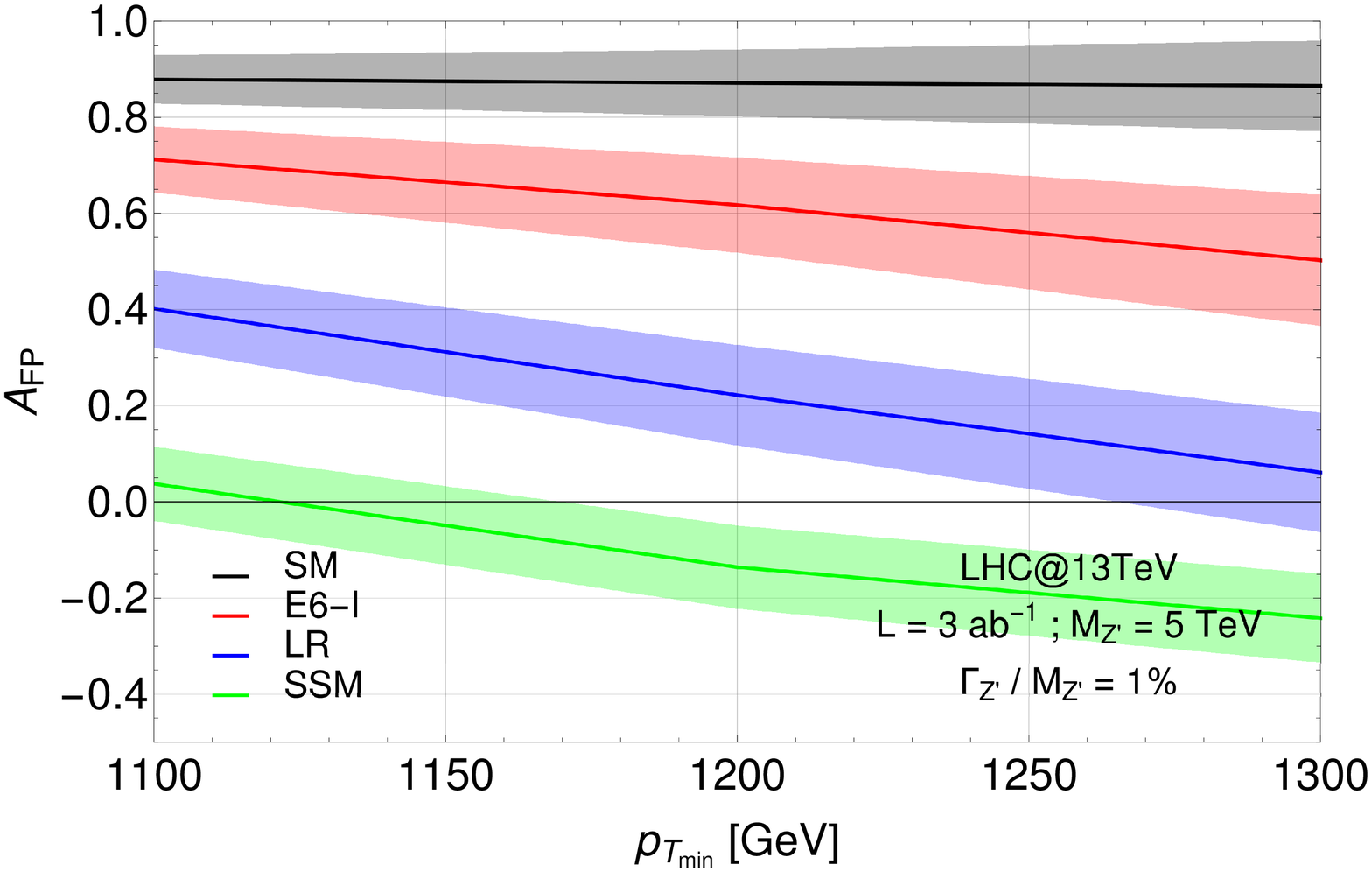}{(a)}
\includegraphics[width=0.45\textwidth]{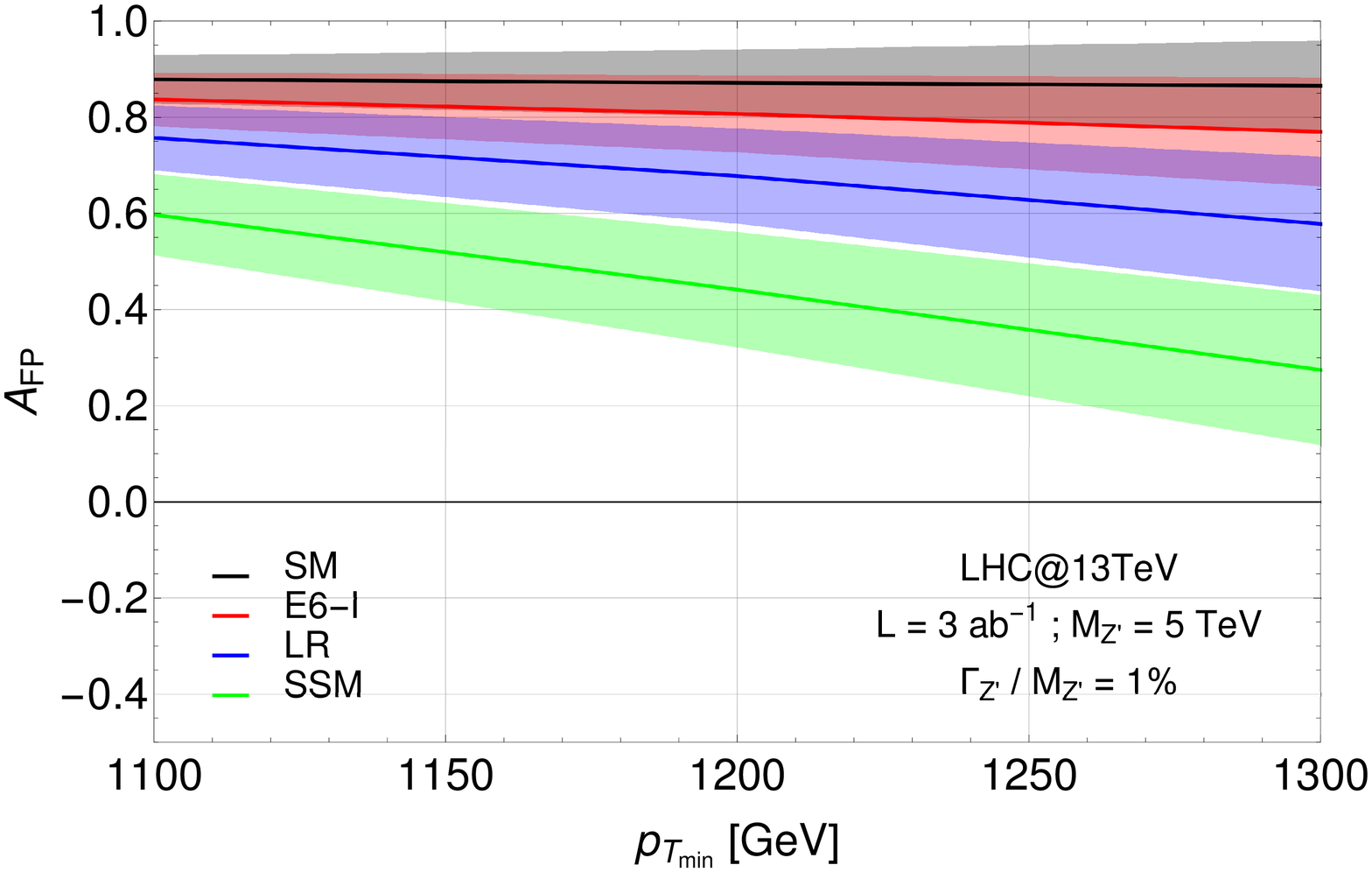}{(b)}
\caption{$A_{\rm FP}$ central value and statistical 1$\sigma$ error band as function of $p_T^{\rm min}$ cut for the LHC at 13 TeV 
and $\mathcal{L}=3~{\rm ab}^{-1}$. The black line represents the SM while the coloured lines represent the three benchmark models.
The mass of the $Z^\prime$ boson is fixed at 5 TeV while its width over mass ratio $\Gamma_{Z^\prime} / M_{Z^\prime}$ has been fixed 
to 1\% (a) and 5\% (b). The values for the FP are chosen in accordance to the tables above.}
\label{fig:AFP_models_5}
\end{center}
\end{figure}

Still, we can exploit the discriminative power of $A_{\rm FP}$ against the SM background and amongst classes of models, ultimately 
extracting constraints that we can impose on the resonance width.
With this is mind, we compare the $A_{\rm FP}$ predictions for the usual three classes of models for different widths, in 
Figs.~\ref{fig:AFP_models_4}--\ref{fig:AFP_models_5}, where we are showing $A_{\rm FP}$ and its statistical error for the three 
$Z^\prime$ benchmarks and SM as a function of $p_T^{\rm min}$ for two values of the resonance mass and different widths. 
As we can see, for a $Z^\prime$ boson mass around 4 TeV, the $A_{\rm FP}$ observable can distinguish between different models 
having $\Gamma_{Z^\prime} / M_{Z^\prime} \sim 10\%$ and in some cases up to 20\% too. 
For a resonance mass around 5 TeV, instead, the sensitivity upon the different classes of models holds up to 
$\Gamma_{Z^\prime} / M_{Z^\prime} \sim 5\%$.

\begin{figure}[h]
\begin{center}
\includegraphics[width=0.45\textwidth]{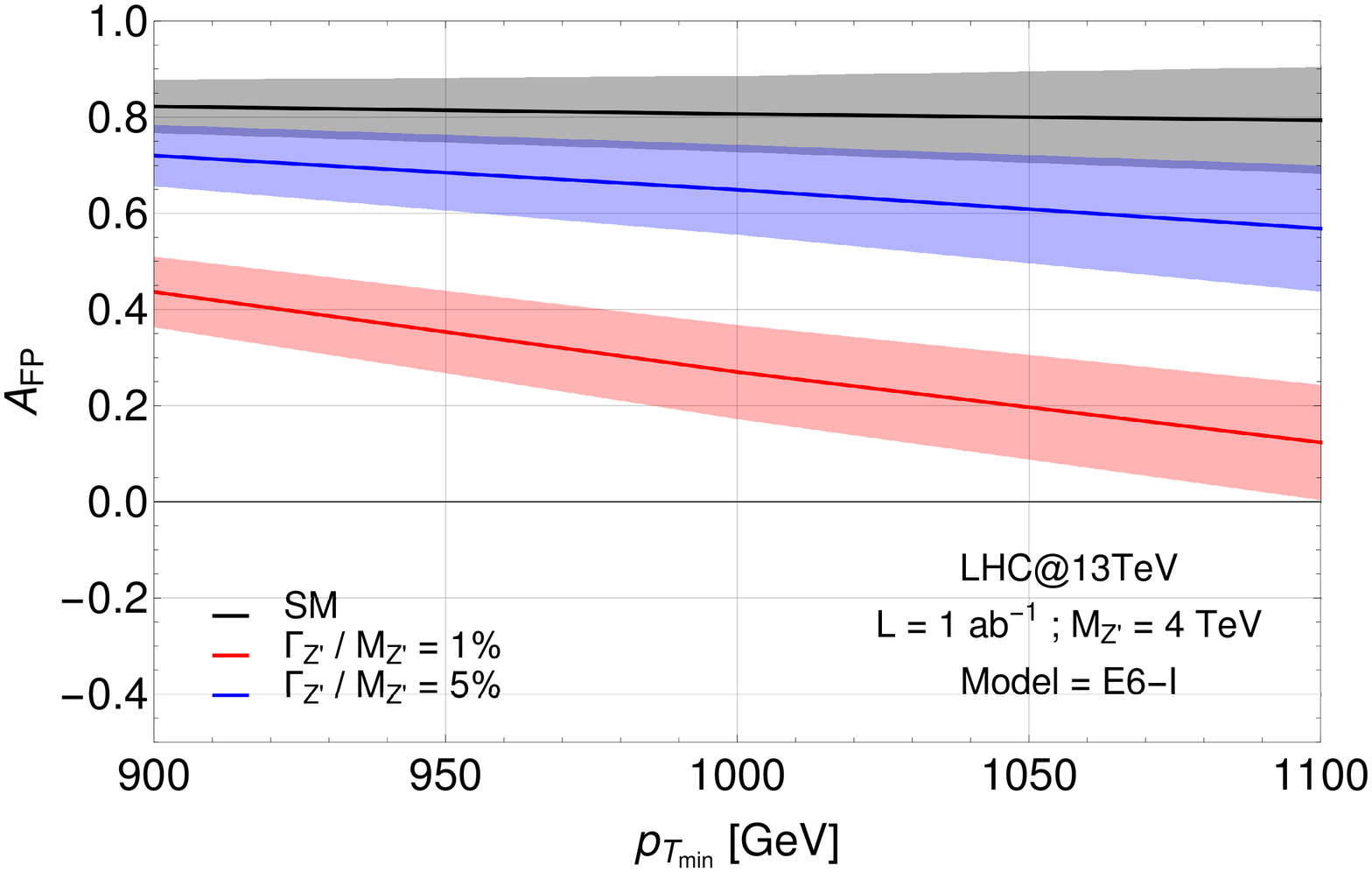}{(a)}
\includegraphics[width=0.45\textwidth]{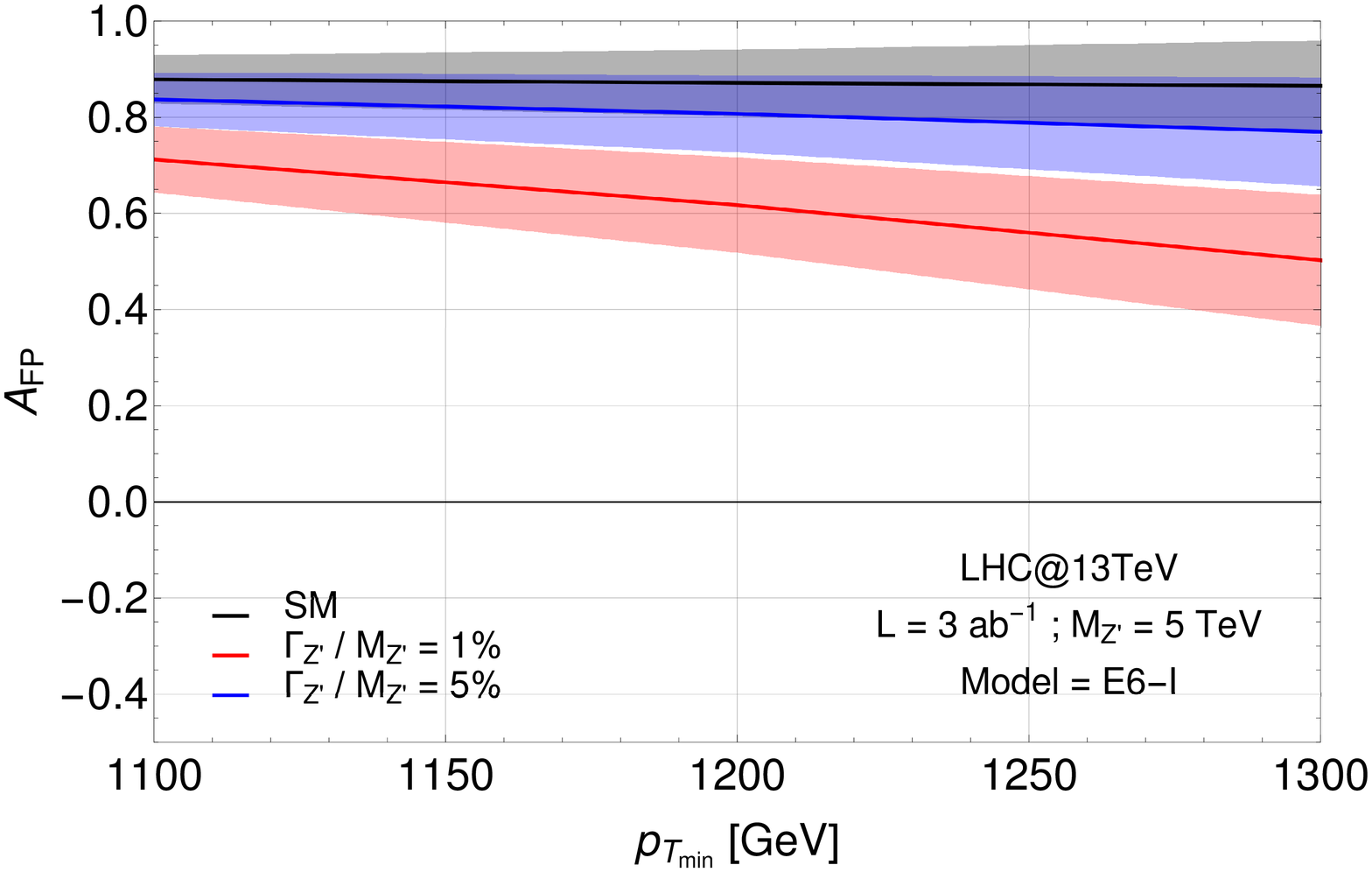}{(b)}
\includegraphics[width=0.45\textwidth]{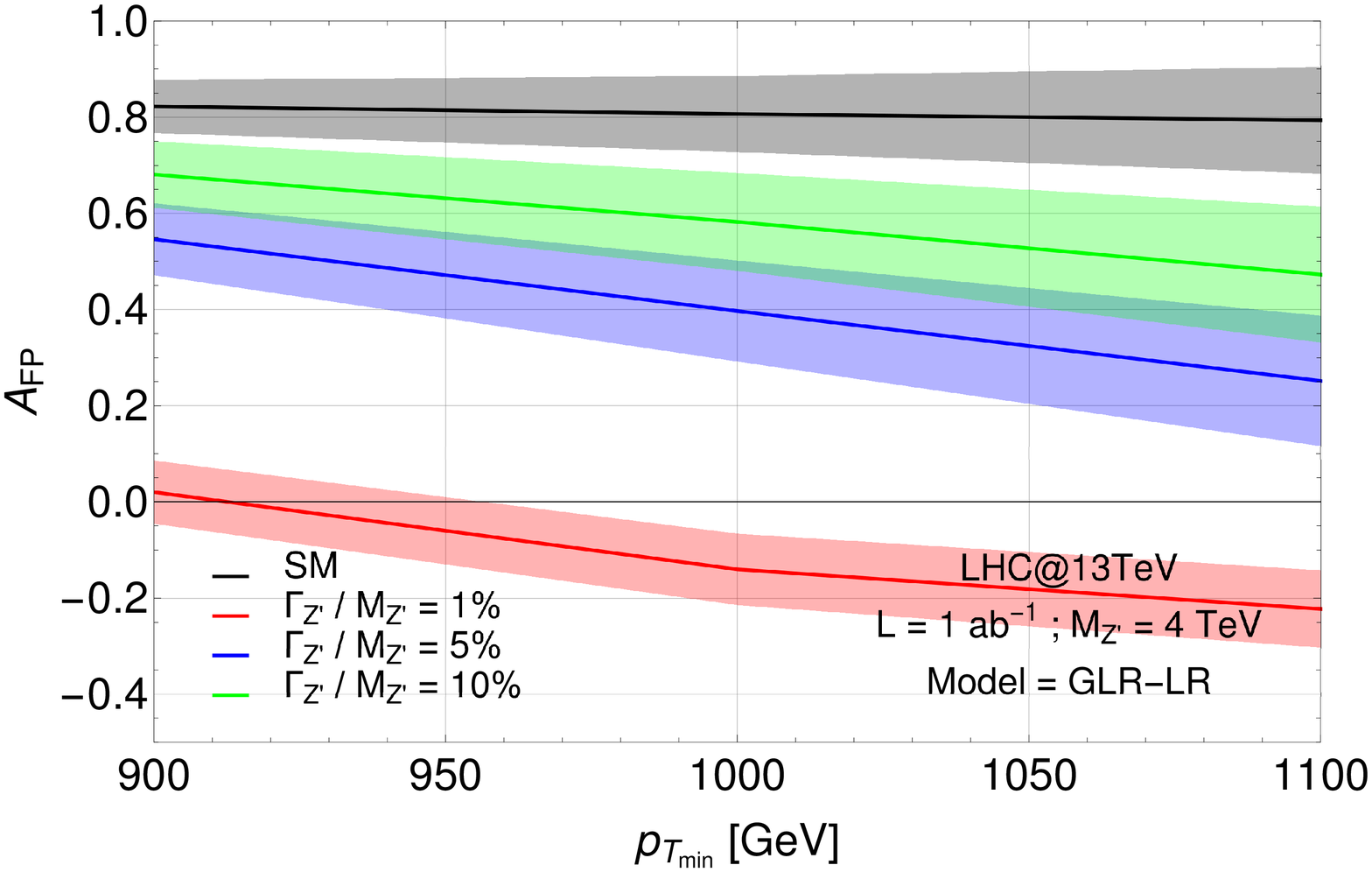}{(c)}
\includegraphics[width=0.45\textwidth]{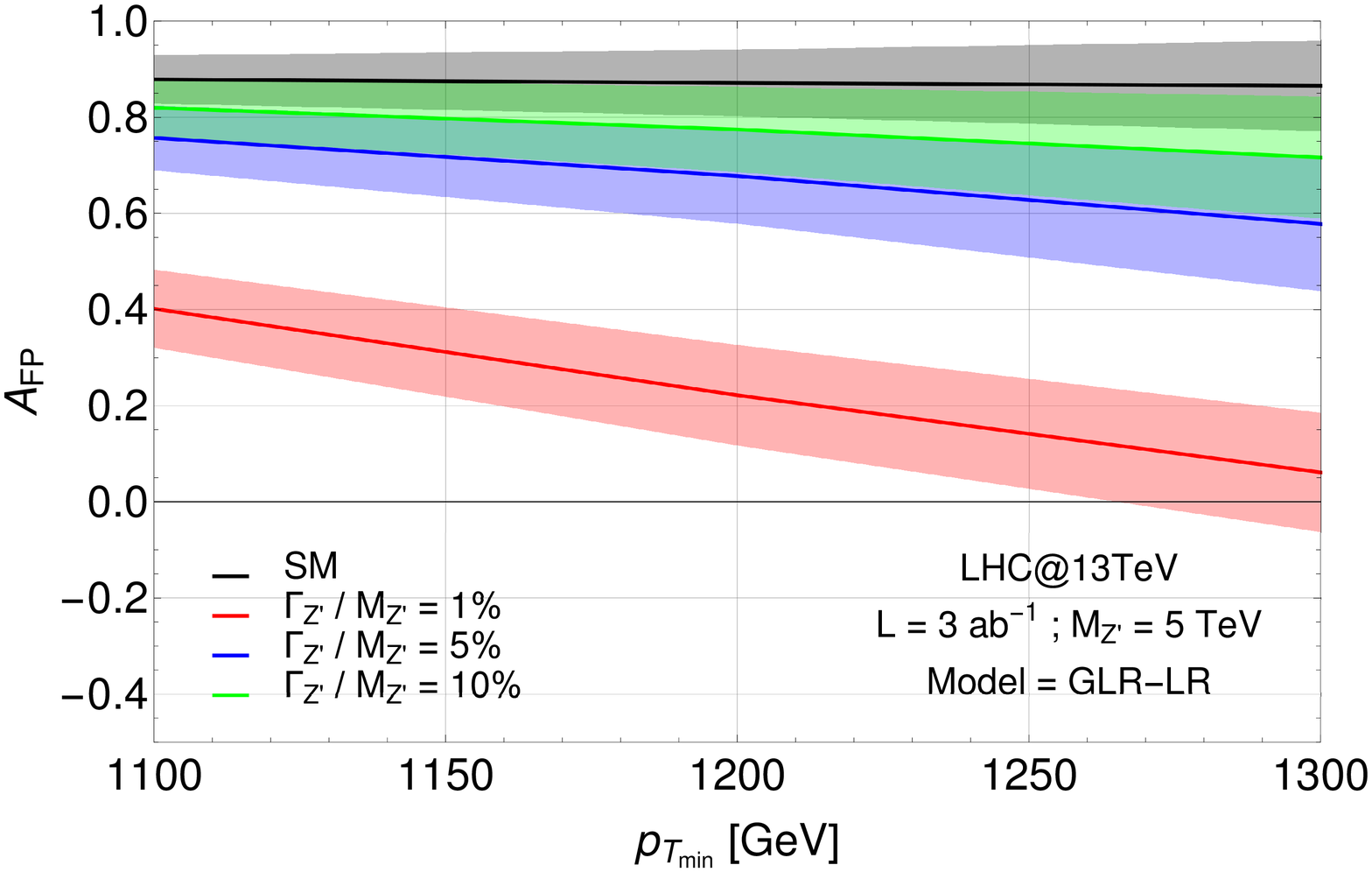}{(d)}
\includegraphics[width=0.45\textwidth]{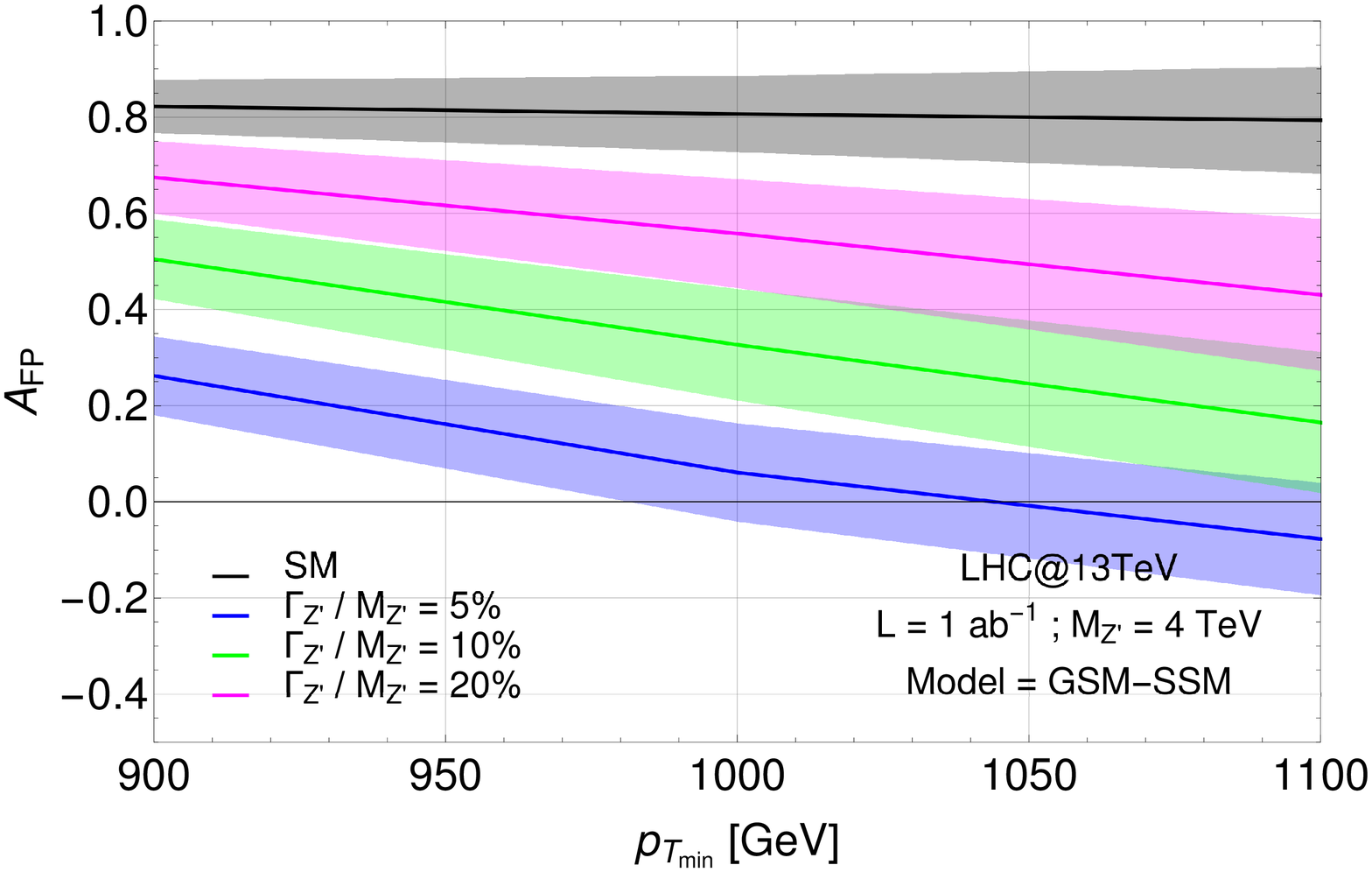}{(e)}
\includegraphics[width=0.45\textwidth]{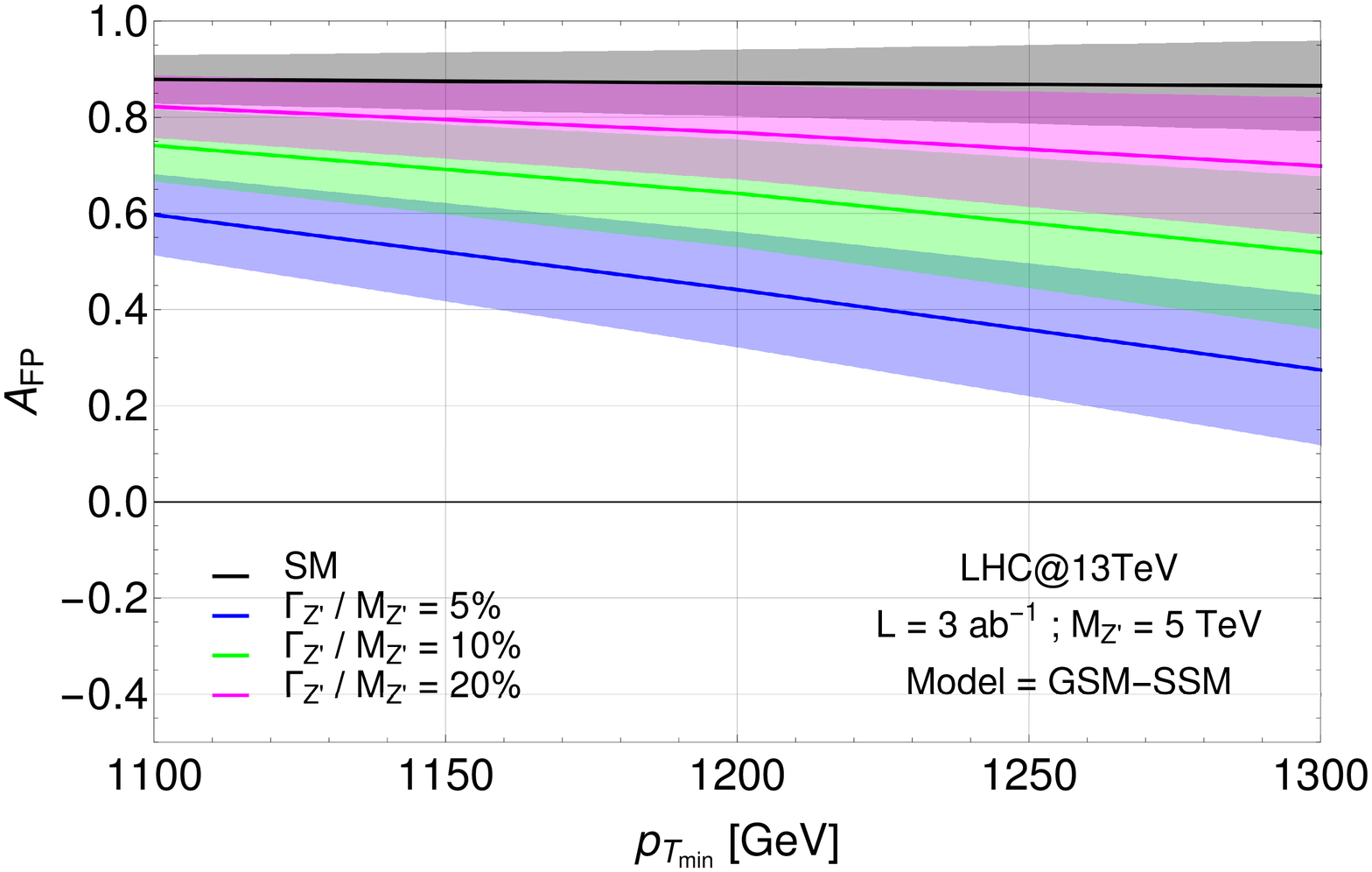}{(f)}
\caption{$A_{\rm FP}$ central value and statistical 1$\sigma$ error band as function of $p_T^{\rm min}$ cut for the LHC at 13 TeV 
and $\mathcal{L}=1~{\rm ab}^{-1}$. The black line represents the SM while the coloured lines represent four different widths (1\%, 5\%, 10\% 
and 20\%) of the $Z^\prime$ resonance in the $E_6^I$ (a), $LR$ (c) and SSM (e) model with a mass of the $Z^\prime$ boson fixed at 4 TeV.
The values for the FP are chosen in accordance to the tables above.
Similarly we repeat the same exercise for the $E_6^I$ (b), $LR$ (d) and SSM (f) model with a mass of the $Z^\prime$ boson fixed at 5 TeV
and and $\mathcal{L}=3~{\rm ab}^{-1}$.}
\label{fig:AFP_widths}
\end{center}
\end{figure}

Finally, coming back to our original purpose, we want to discuss now the sensitivity of $A_{\rm FP}$ upon the resonance width.
In Figs.~\ref{fig:AFP_widths} we are showing its discriminative power against the resonance width within each class for two choices 
of the $Z^\prime$ boson mass. The $A_{\rm FP}$ observable seems to fulfil the task: within each class of models we are able to set 
important constraints on the resonance width. In the case of resonances of the order of 4 TeV, assuming an integrated luminosity of 1 ab$^{-1}$, 
we would be able to constrain their widths up to $\Gamma_{Z^\prime} / M_{Z^\prime} \sim 5\% $ within the $E_6$ class of models, 
up to $\Gamma_{Z^\prime} / M_{Z^\prime} \sim 10\% $ within the $LR$ class of models and 
up to $\Gamma_{Z^\prime} / M_{Z^\prime} \sim 20\% $ within the $SSM$ class of models
For resonances of the order of 5 TeV we obtain similar results, assuming an integrated luminosity of 3 ab$^{-1}$.

\section{Conclusions}
\label{sec:summa}

In summary, we have defined a new kinematic asymmetry, $A_{\rm FP}$,
based around a FP appearing in the normalised transverse momentum
distribution of either lepton in DY processes. The remarkable features
of this FP are its insensitivity to the underlying $Z^\prime$ model as
well as quantities which carry (theoretical) systematic errors such as
PDFs and their factorisation and renormalisation scales. Hence, this FP
displays model-independent characteristics, as it is only sensitive to
the collider energy (which is known) and the mass of the intervening
$Z^\prime$ (which is expected to be extracted from the di-lepton
invariant mass).

In fact, while the FP location is stable
against variations of the $Z^\prime$ boson width, the $A_{\rm FP}$
asymmetry strongly dependent upon the width. 
The combination of these features makes of $A_{\rm FP}$ a suitable
observable to determine the characteristics of any $Z^\prime$ which
may be discovered at the LHC. 
Lastly, the $A_{\rm FP}$ could also be used to limit the possible
range of widths of a $Z^\prime$ signal which could be used as a
constraint in a fit of a resonance peak in an invariant mass spectrum.

Finally, we remark that the effectiveness of the new variable
will increase significantly with the LHC luminosity, so as to expect that its importance
will be appreciated after a few years of Run 2 (i.e., after some 300 fb$^{-1}$ of data)
or else rather immediately at a future High-Luminosity LHC (HL-LHC) stage~\cite{Gianotti:2002xx}
(i.e., starting from  1 ab$^{-1}$ of data), depending on the $Z^\prime$ mass, width and couplings.

\section*{Acknowledgements}
\noindent
This work is supported by the Science and Technology Facilities Council, grant number ST/L000296/1.
All authors acknowledge partial financial support through the NExT Institute.

\bibliography{bib}

\end{document}